\documentclass[12pt]{article}
\pdfoutput=1

\usepackage{amsmath,amssymb,amscd}
\usepackage{listings}
\usepackage{caption}
\usepackage{dsfont}
\usepackage{slashed}
\usepackage{color}

\usepackage[pdftex]{graphicx}
\usepackage{epstopdf}
\usepackage{subfigure}
\usepackage{epsfig}
\usepackage{listings}
\usepackage{caption}
\usepackage{cite}
\usepackage{fontawesome}

\usepackage{url}

\usepackage{multirow}

\newcommand{\bea}{\begin{align}}
\newcommand{\eea}{\end{align}}
\newcommand{\beq}{\begin{equation}}
\newcommand{\eeq}{\end{equation}}
\newcommand{\nbea}{\begin{align*}}
\newcommand{\neea}{\end{align*}}
\newcommand{\nbeq}{\begin{equation*}}
\newcommand{\neeq}{\end{equation*}}

\newcommand{\ket}[1]{\left| #1 \right>} % for Dirac bras
\newcommand{\bra}[1]{\left< #1 \right|} % for Dirac kets
\newcommand{\braket}[2]{\left< #1 \vphantom{#2} \right|
 \left. #2 \vphantom{#1} \right>} % for Dirac brackets

\newcommand{\Op}{\mathbb{O}}
        
\numberwithin{equation}{section}

\usepackage{tabularx}

\newcommand\T{\rule{0pt}{2.8ex}}
\newcommand\B{\rule[-1.8ex]{0pt}{0pt}}

\usepackage{multicol}
\usepackage{color}
\definecolor{verdes}{cmyk}{0.92,0,0.59,0.4}

\definecolor{Grn}{rgb}{0.1,0.5,0.2}
\definecolor{Blu}{rgb}{0.,0.,1.}
\definecolor{Red}{rgb}{0.7,0.1,0.1}
\definecolor{SE}{rgb}{0.5,0,0.4}

\newcommand{\Red}[1]{{\color{Red}{#1}}}

\newcommand{\niceRed}[1]{\Red{#1}}

\usepackage{comment}

\usepackage{hyperref} 
\hypersetup{
     colorlinks = true,
     linkcolor = blue,
     anchorcolor = blue,
     citecolor = magenta,
     filecolor = blue,
     urlcolor = magenta,
     bookmarks=true,
     linktocpage=true
     }

\begin{document}

\begin{titlepage}

\baselineskip=21pt
 {\small
 \rightline{CERN-TH-2020-104}
 \rightline{CALT-TH-2020-027}
 }
\vskip 0.75in

\begin{center}

{\large {\bf The Fermionic Universal One-Loop Effective Action}}

\vskip 0.3in

{\small
{\bf Sebastian~A.~R.~Ellis}$^{1}$,
{\bf J\'er\'emie~Quevillon}$^{2}$,
{\bf Pham~Ngoc~Hoa~Vuong}$^{2}$,\\
{\bf Tevong~You}$^{3,4,5}$
and {\bf Zhengkang~Zhang}$^{6}$
}

{\small {\it
\vspace{0.75cm}
$^1${SLAC National Accelerator Laboratory, 2575 Sand Hill Road, Menlo Park CA 94025, USA}\\[4pt]
$^2${Laboratoire de Physique Subatomique et de Cosmologie,\\ Universit\'{e} Grenoble-Alpes, CNRS/IN2P3, Grenoble INP, 38000 Grenoble, France} \\[4pt]
$^3${CERN, Theoretical Physics Department, Geneva, Switzerland}\\[4pt]
$^4${DAMTP, University of Cambridge, Wilberforce Road, Cambridge, CB3 0WA, UK} \\[4pt]
$^5${Cavendish Laboratory, University of Cambridge, J.J. Thomson Avenue, Cambridge, CB3 0HE, UK 
} \\[4pt]
$^6${Walter Burke Institute for Theoretical Physics, \\California Institute of Technology, Pasadena, CA 91125, USA} \\
}}

\vskip 0.4in

{\bf Abstract}

\end{center}

\baselineskip=18pt \noindent

%%%%%%%%%%%%%%%%%%%%%%%%%%%%%%%%%%%%%%%%%%%%%%%%%

{\small

Recent development of path integral matching techniques based on the covariant derivative expansion has made manifest a universal structure of one-loop effective Lagrangians. The universal terms can be computed once and for all to serve as a reference for one-loop matching calculations and to ease their automation. 
Here we present the fermionic universal one-loop effective action (UOLEA), resulting from integrating out heavy fermions with scalar, pseudo-scalar, vector and axial-vector couplings. We also clarify the relation of the new terms computed here to terms previously computed in the literature and those that remain to complete the UOLEA. Our results can be readily used to efficiently obtain analytical expressions for effective operators arising from heavy fermion loops. \href{https://github.com/HoaVuong-lpsc/The-Fermionic-UOLEA-Mathematica-notebook}{\faGithub}

}

%%%%%%%%%%%%%%%%%%%%%%%%%%%%%%%%%%%%%%%%%%%%%%%%

\end{titlepage}

\tableofcontents

%*******************************************************************************************
\section{Introduction and summary of results}

The methods of effective field theory (EFT) have seen a resurgence lately in particle physics, due in part to the lack of new physics discovery at the weak scale. If new physics is indeed decoupled to heavier scales, as observations seem to be indicating, then the Standard Model (SM) should be properly considered as an EFT supplemented by higher-dimensional operators. The coefficients of these higher-dimensional operators encapsulate the new physics integrated out at some higher energy scale. Calculating these coefficients from ultraviolet (UV) theories has traditionally been performed using Feynman diagrams, where amplitudes involving the heavy degrees of freedom are explicitly ``matched'' to the EFT amplitudes. However, a more elegant approach is to ``integrate out'' the heavy particles by evaluating the path integral directly~\cite{Gaillard:1985uh,Chan:1986jq,Cheyette:1987qz,Henning:2014wua, Drozd:2015rsp, Henning:2016lyp, Ellis:2016enq, Fuentes-Martin:2016uol, Zhang:2016pja, Ellis:2017jns, Kramer:2019fwz, Cohen:2019btp}. While the adoption of this approach for practical phenomenological calculations has been limited in the past by cumbersome expansion techniques and the misconception that it could not account for matching with both heavy and light particles in the loop, these technical issues have been addressed in the last few years~\cite{Henning:2016lyp,Ellis:2016enq,Fuentes-Martin:2016uol,Zhang:2016pja}. New methods were developed to evaluate the path integral at one loop more efficiently using improved expansion techniques (as for example the covariant diagram method~\cite{Zhang:2016pja}), that could also include mixed heavy-light matching.

Compared to the traditional approach of matching Feynman diagrams, these path integral methods have several advantages: they can be calculated more generally, directly and systematically when computing a set of operator coefficients. Ultimately, it was pointed out in Refs.~\cite{Henning:2014wua, Drozd:2015rsp} that the one-loop effective action has a universal structure which makes repeated evaluation of the path integral redundant. It is this set of universal terms and coefficients, evaluated once and for all, that forms the so-called Universal One-Loop Effective Action (UOLEA). Starting from the UOLEA, a one-loop matching calculation is reduced to an algebraic manipulation of matrix traces.  

The piece of the UOLEA that was first worked out, for the simplified case of degenerate masses in Ref.~\cite{Henning:2014wua} and generalised to the non-degenerate case in Ref.~\cite{Drozd:2015rsp}, contains terms arising from integrating out heavy bosonic fields $\Phi$ which couple to light fields $\phi$ via a Lagrangian of the form
\begin{equation}
\mathcal{L}_\text{UV}[\Phi, \phi] = \mathcal{L}_{0}[\phi] 
+ \Phi^\dagger\left(P^2 - M^2 - U[\phi]\right)\Phi + \mathcal{O}(\Phi^3) \, ,
\label{eq:LUVbosonic}
\end{equation}
where $P_\mu \equiv i D_\mu$ is the Hermitian covariant derivative, $M$ is a diagonal mass matrix for the heavy fields $\Phi$, and the model-dependent couplings of $\Phi$ to $\phi$ are encapsulated in the matrix $U[\phi]$. By virtue of keeping the covariant derivatives intact, the UOLEA can thus be written as an expansion in covariant derivatives, i.e.\ a covariant derivative expansion (CDE)~\cite{Gaillard:1985uh,Chan:1986jq,Cheyette:1987qz}. In the end, to obtain the low-energy EFT Lagrangian up to dimension-six operators, one simply needs to insert the matrix $U[\phi]$ into the UOLEA: 
\begin{align}
\mathcal{L}_\text{UOLEA}^\text{bosonic,heavy} 
&=-i c_s \,\mathrm{tr}\, \Bigl\{ 
f_2^i\, {U}_{ii} 
+f_3^i\, G^{\prime\mu\nu}_i G^\prime_{\mu\nu,i} 
+f_4^{ij}\, {U}_{ij} {U}_{ji} \nonumber\\ 
&\qquad\quad
+f_5^i\, [P^\mu, G^\prime_{\mu\nu,i}] [P_\rho, G^{\prime\rho\nu}_i]
+f_6^i\, G^{\prime\mu}_{\;\;\,\nu,i} G^{\prime\nu}_{\;\;\,\rho,i} G^{\prime\rho}_{\;\;\,\mu,i} \nonumber\\
&\qquad\quad
+f_7^{ij}\, [P^\mu, {U}_{ij}] [P_\mu, {U}_{ji}]
+f_8^{ijk}\, {U}_{ij} {U}_{jk} {U}_{ki}
+f_9^i\, {U}_{ii} G^{\prime\mu\nu}_i G^\prime_{\mu\nu,i} \nonumber\\ 
&\qquad\quad
+f_{10}^{ijkl}\, {U}_{ij} {U}_{jk} {U}_{kl} {U}_{li}
+f_{11}^{ijk}\, {U}_{ij} [P^\mu, {U}_{jk}] [P_\mu, {U}_{ki}] \nonumber\\ 
&\qquad\quad
+f_{12}^{ij}\, \bigl[P^\mu, [P_\mu, {U}_{ij}]\bigr] \bigl[P^\nu, [P_\nu, {U}_{ji}]\bigr]
+f_{13}^{ij}\, {U}_{ij} {U}_{ji} G^{\prime\mu\nu}_i G^\prime_{\mu\nu,i} \nonumber\\ 
&\qquad\quad
+f_{14}^{ij}\, [P^\mu, {U}_{ij}] [P^\nu, {U}_{ji}] G^\prime_{\nu\mu, i}
\nonumber\\
&\qquad\quad
+f_{15}^{ij}\, \bigl( {U}_{ij} [P^\mu, {U}_{ji}] -[P^\mu, {U}_{ij}] {U}_{ji} \bigr) [P^\nu, G^\prime_{\nu\mu,i}] \nonumber\\ 
&\qquad\quad
+f_{16}^{ijklm}\, {U}_{ij} {U}_{jk} {U}_{kl} {U}_{lm} {U}_{mi} \nonumber\\ 
&\qquad\quad
+f_{17}^{ijkl}\, {U}_{ij} {U}_{jk} [P^\mu, {U}_{kl}] [P_\mu, {U}_{li}]
+f_{18}^{ijkl}\, {U}_{ij} [P^\mu, {U}_{jk}] {U}_{kl} [P_\mu, {U}_{li}] \nonumber\\ 
&\qquad\quad
+f_{19}^{ijklmn}\, {U}_{ij} {U}_{jk} {U}_{kl} {U}_{lm} {U}_{mn} {U}_{ni}
\Bigr\} 
\label{eq:heavyonlybosonicUOLEA}\\
&= \sum_N f_N^{(P)} \mathbb{O}_N^{(P)} + f_N^{(U)} \mathbb{O}_N^{(U)} + f_N^{(PU)} \mathbb{O}_N^{(PU)} \,.
\label{eq:heavyonlybosonicUOLEA0}
\end{align}
The prefactor $c_s = \frac{1}{2}$ for each real degree of freedom (e.g.\ real scalar, vector) and can be taken as $c_s = \pm 1$ in some other cases~\cite{Henning:2014wua}. The {\it universal coefficients} $f_N^{ij\dots}$ are functions of heavy particle masses $m_i, m_j,\dots$, and are expressed in terms of a set of master integrals. The field strength matrix is defined as $G^{\prime}_{\mu\nu} = -[P_\mu,\,P_\nu]=-igG_{\mu\nu}$, and the subscripts $i,j,\dots$ on $G$ and $U$ instruct us to take the corresponding block for particles $i,j,\dots$. In Eq.~\eqref{eq:heavyonlybosonicUOLEA0}, we have schematically summarised the entire expression by three UOLEA operator classes: those involving only covariant derivatives ($\mathbb{O}_N^{(P)}$), only interaction matrices ($\mathbb{O}_N^{(U)}$), and both ($\mathbb{O}_N^{(PU)}$). We refer the reader to Refs.~\cite{Henning:2014wua,Drozd:2015rsp,Zhang:2016pja} for the derivation of this bosonic UOLEA, though we stress that it is no longer necessary to re-do the path integral calculation for each specific model given the availability of these universal results. The UOLEA operator structures, written in terms of the matrices $P$ and $U$, become EFT operators when substituting in the specific forms of these matrices, in terms of the light fields and for a given UV model, which can then be rearranged into the desired non-redundant EFT basis.\footnote{Note that the UOLEA can be expanded indefinitely in the CDE; in Eq.~\eqref{eq:heavyonlybosonicUOLEA} and, later, the fermionic UOLEA we terminate the CDE to keep only all UOLEA operator structures necessary for obtaining EFT operators up to dimension six.}

There are, however, additional structures that arise in some UV Lagrangians which lead to new terms in the UOLEA beyond those in Eq.~\eqref{eq:heavyonlybosonicUOLEA}. In particular, for UV Lagrangians containing heavy fermion fields, further terms in the UOLEA arise from fermionic loops. While some of them can be obtained from the bosonic UOLEA \eqref{eq:heavyonlybosonicUOLEA} by ``squaring'' the functional determinant (see e.g.\ Appendix~A1 of Ref.~\cite{Henning:2014wua} and Appendix~E of Ref.~\cite{Drozd:2015rsp}) to put the UV Lagrangian into the form of Eq.~\eqref{eq:LUVbosonic}, this only yields partial results when the interactions involve $\gamma$ matrices. It is therefore necessary to extend the UOLEA to properly include fermionic loops.

In this work, we present this fermionic UOLEA. It can be applied straightforwardly to the case of fermions in an analogous manner to the bosonic case described above. Specifically, we consider a UV Lagrangian for a heavy multiplet of fermions $\Psi$ interacting with a light multiplet of bosons or fermions $\phi$ of the following form:
\begin{align}
\mathcal{L}_{\rm UV}[\phi, \Psi] = \mathcal{L}_0[\phi] +  \bar{\Psi} \left( \slashed{P} - M - X[\phi] \, \right) \Psi\, ,
\label{eq:LUVfermionic}
\end{align}
and we decompose the interaction matrix $X[\phi]$ into scalar, pseudoscalar, vector, and axial-vector coupling matrices as
\begin{align}
X[\phi] = W_0[\phi] +  W_1[\phi]\,i\gamma^5 + V_{\mu}[\phi]\gamma^{\mu} + A_{\mu}[\phi]\gamma^{\mu}\gamma^5 \, .
\label{eq:Xmatrix}
\end{align}
The low-energy EFT at one loop is then obtained by substituting these matrices into our fermionic UOLEA, which reads schematically
\begin{align}\label{eq:UOLEA_fermionic}
\mathcal{L}_\text{UOLEA}^\text{fermionic,heavy} =&\; \sum_{N} 
f_N^{(P)} \Op_N^{(P)} + 
f_N^{(W_0)} \Op_N^{(W_0)} + 
f_N^{(W_1)} \Op_N^{(W_1)} +
f_N^{(W_0 W_1)} \Op_N^{(W_0 W_1)} 
\nonumber \\[-4pt]
&\qquad +
f_N^{(P W_0)} \Op_N^{(P W_0)} +
f_N^{(P W_1)} \Op_N^{(P W_1)} +
f_N^{(P W_0 W_1)} \Op_N^{(P W_0 W_1)} \nonumber\\[4pt]
&\qquad + 
f_N^{(V)} \Op_N^{(V)} + 
f_N^{(A)} \Op_N^{(A)} +
f^{(VA)}_N \Op_N^{(VA)} 
\nonumber\\
&\qquad +
f^{(PV)}_N \Op_N^{(PV)} +
f^{(PA)}_N \Op_N^{(PA)} +
f^{(PVA)}_N \Op_N^{(PVA)} 
\nonumber\\
&\qquad + 
f_N^{(W_0V)} \Op_N^{(W_0V)} +
f_N^{(W_1V)} \Op_N^{(W_1V)} +
f_N^{(W_0W_1V)} \Op_N^{(W_0W_1V)} \nonumber\\
&\qquad +
f_N^{(PW_0V)} \Op_N^{(PW_0V)} +
f_N^{(PW_1V)} \Op_N^{(PW_1V)} +
f_N^{(PW_0W_1V)} \Op_N^{(PW_0W_1V)} \nonumber\\
&\qquad +
f_N^{(W_0A)} \Op_N^{(W_0A)} +
f_N^{(W_1A)} \Op_N^{(W_1A)} +
f_N^{(W_0W_1A)} \Op_N^{(W_0W_1A)} \nonumber\\
&\qquad +
f_N^{(PW_0A)} \Op_N^{(PW_0A)} +
f_N^{(PW_1A)} \Op_N^{(PW_1A)} +
f_N^{(PW_0W_1A)} \Op_N^{(PW_0W_1A)} \nonumber\\
&\qquad +
f_N^{(W_0VA)} \Op_N^{(W_0VA)} +
f_N^{(W_1VA)} \Op_N^{(W_1VA)} +
f_N^{(W_0W_1VA)} \Op_N^{(W_0W_1VA)} \nonumber\\
&\qquad +
f_N^{(PW_0VA)} \Op_N^{(PW_0VA)} +
f_N^{(PW_1VA)} \Op_N^{(PW_1VA)} +
f_N^{(PW_0W_1VA)} \Op_N^{(PW_0W_1VA)} 
\, .
\end{align}
There are a large combinatorial number of possibilities for the fermionic UOLEA operator structures when including all coupling matrices in Eq.~\eqref{eq:Xmatrix}. We will see, however, that when calculating specific cases one can employ power counting to pick out the UOLEA operator structures that are relevant for matching to a set of desired EFT operators. Moreover, if the low-energy EFT does not contain massive vector bosons (e.g.\ arising from a broken gauge symmetry), then only the first two lines of Eq.~\eqref{eq:UOLEA_fermionic} are needed, comprising a relatively compact set of UOLEA operators. These UOLEA operator structures, along with their universal coefficients for the degenerate mass case, are tabulated below in Tables~\ref{tab:pureP},~\ref{tab:mixedPW0},~\ref{tab:mixedPW1} and~\ref{tab:mixedPW0W1}. Explicit results for the non-degenerate case and for the rest of Eq.~\eqref{eq:UOLEA_fermionic} are available in a Mathematica notebook on \texttt{GitHub}~\href{https://github.com/HoaVuong-lpsc/The-Fermionic-UOLEA-Mathematica-notebook}{\faGithub},~\cite{notebookUOLEA}, as explained in Sec.~\ref{sec:ResultsDescription}.

\begin{table}[t]
\begin{center}
\begin{tabular}{|c|ccc|}
\multicolumn{4}{c}{\T\B \niceRed{\textbf{Universal terms available in the UOLEA}} }  \\
\hline 
\T\B  & Heavy-only  & Mixed heavy-light & + derivative couplings \\
\hline
\T\B Bosonic & $\checkmark$ \cite{Drozd:2015rsp}  & $\checkmark$ \cite{Ellis:2017jns} & $-$ \\
\T\B Fermionic & $\qquad\checkmark$ [this work]~\href{https://github.com/HoaVuong-lpsc/The-Fermionic-UOLEA-Mathematica-notebook}{\faGithub}  & $(\checkmark)$  & $\quad-^{(*)}$ \\
\T\B Mixed statistics & $(\checkmark)$  & $(\checkmark)$ & $\quad-^{(*)}$  \\
\hline
\multicolumn{4}{l}{\footnotesize $^{(*)}$ do not arise in renormalizable UV theories.\rule{0pt}{3.5ex}} \\
\end{tabular}
\caption{\label{tab:UOLEAprogress}
Status of the UOLEA. 
Entries marked by ``$\checkmark$'' are available in the form of operator structures built from the various types of couplings that appear in the quadratic Lagrangian. Entries marked by ``$(\checkmark)$'' are not available in the same form, but can be computed by plugging fermion couplings into the results of Ref.~\cite{Kramer:2019fwz} and evaluating Dirac matrix traces. Entries marked by ``$-$'' have not been computed in the literature, though the techniques for computing them are available. See text for details.
}
\end{center}
\end{table}

The bosonic UOLEA presented in Ref.~\cite{Drozd:2015rsp} (summarised above in Eq.~\eqref{eq:heavyonlybosonicUOLEA0}) and fermionic UOLEA presented in this paper (summarised above in Eq.~\eqref{eq:UOLEA_fermionic}) complete the one-loop matching master formula that includes loops involving heavy bosonic fields and heavy fermionic fields, respectively, for UV theories whose Lagrangians take the form of Eq.~\eqref{eq:LUVbosonic} or Eq.~\eqref{eq:LUVfermionic}, and for up to dimension-6 operators in the EFT. Other UV theories exhibit additional coupling structures which are not captured by these UOLEAs, such as tensor current coupling (involving $\sigma_{\mu\nu}$), derivative couplings (which give rise to ``open covariant derivatives'' in the quadratic Lagrangian) and mixed bosonic-fermionic loops. If the UV Lagrangian includes terms coupling heavy fields linearly to the light fields, $\mathcal{L}_{\rm UV} \supset \Phi^\dagger F[\phi] + \text{h.c.}$, then mixed heavy-light loops also contribute.\footnote{Linear couplings also generate tree-level contributions, but loop-level mixed heavy-light contributions can in certain cases be the leading terms for certain operators~\cite{Jiang:2018pbd}.} For the bosonic case, the mixed heavy-light terms, $\mathcal{L}_\text{UOLEA}^\text{bosonic,mixed}$, were computed in Ref.~\cite{Ellis:2017jns}, where it was found that the operator structures in $\mathcal{L}_\text{UOLEA}^\text{bosonic,mixed}$ mirror those in $\mathcal{L}_\text{UOLEA}^\text{bosonic,heavy}$ with a much larger number of terms due to the heavy-light combinatorics. We expect the same for the fermionic UOLEA, though given the proliferation of terms already in the heavy-only case we find it less compelling to also tabulate the mixed heavy-light terms explicitly. 

In Table~\ref{tab:UOLEAprogress} we summarise the progress of the UOLEA program. Fermionic results are also available in Ref.~\cite{Kramer:2019fwz}, though not in the same form as our expressions since the various matrix substructures are not expanded as in Eq.~\eqref{eq:Xmatrix} --- they can be computed, together with additional structures such as mixed fermion-boson and heavy-light loops, after plugging in these substructures and further evaluating the resulting Dirac matrix traces. Finally, UV theories involving derivative couplings generate additional terms in the UOLEA which have not yet been computed, though for the fermionic case they only arise when matching to non-renormalisable UV Lagrangians.
The UOLEA has, so far, also been limited to those terms necessary for obtaining EFT operators up to dimension six only. Nevertheless, it is worth emphasising that, following the technical development of evaluating one-loop functional determinants with general structures~\cite{Henning:2016lyp,Fuentes-Martin:2016uol,Zhang:2016pja,Cohen:2019btp}, {\it one-loop functional matching is a fully solved problem}, independently of the availability of the UOLEA that captures those additional structures. The usefulness of the UOLEA lies in its packaging of certain universal steps of the calculation into the form of a master formula. 

The paper is organised as follows. In Sec.~\ref{sec:fermionicUOLEA}, we describe our calculation of the fermionic UOLEA and present the final results for the universal coefficients of the UOLEA operators. We then present examples illustrating the use of the fermionic UOLEA for efficient one-loop matching calculations in Sec.~\ref{sec:examples}, before concluding in Sec.~\ref{sec:conclusion}.

\section{The Fermionic UOLEA}
\label{sec:fermionicUOLEA}

Fermions, by virtue of their symmetry properties, necessitate additional care as compared with spin-0 bosons, which have been the primary focus of CDE developments thus far. Some previous work on using the CDE to integrate out heavy fermions had employed the approach of squaring the argument of the functional trace in the effective action so as to bring it into the same form as bosonic loops, for subsequent insertion into the bosonic UOLEA as written in Eq.~\eqref{eq:heavyonlybosonicUOLEA}~\cite{Henning:2014wua,Huo:2015exa}. However, this approach cannot be straightforwardly applied to the case where fermion coupling structures contain gamma matrices beyond that accompanying the covariant derivative $\slashed{P}$.

As was pointed out for example in Refs.~\cite{Henning:2016lyp,Zhang:2016pja}, the argument of the functional trace need not be squared, in which case a CDE and universal one-loop action may be still be formulated, with a somewhat different structure from the bosonic UOLEA of the previous section but one that simplifies the UOLEA as applied to fermions. This procedure was employed in Ref.~\cite{Kramer:2019fwz} to obtain contributions to the UOLEA from integrating out heavy fermions, though they do not decompose the general coupling matrix $X$ into its Hermitian matrix substructure constituents so that their final result still requires taking the trace over $\gamma$ matrices.

Here we provide a master formula in terms of these matrix substructures. In this case, as will be expanded upon in detail in the rest of this section, it is straightforward to account for all possible Lorentz structures of fermionic coupling matrices to light fields, thereby allowing for the completion of a \textit{fermionic UOLEA}.

\subsection{One-loop matching from the path integral}

Let us begin by reviewing the basic idea of one-loop functional matching, focusing on the case of integrating out heavy fermions. Consider a UV Lagrangian containing a multiplet of heavy Dirac fermion fields $\Psi$ and light fields $\phi$. Assuming the heavy fermions $\Psi$ couple to the light fields only via bilinears, the UV Lagrangian can be written in the form
\begin{equation}
\mathcal{L}_{\rm UV}[\phi, \Psi] = \mathcal{L}_0[\phi] +  \bar{\Psi} \left( \slashed{P} - M - X[\phi] \, \right) \Psi\, , 
\label{eq:fermionicUVL}
\end{equation}
where $P_{\mu} \equiv i D_{\mu}$ and $M$ is the diagonal mass matrix for the multiplet $\Psi$. In order to maximise the analytical and physical utility of the universal structures obtained by using the CDE method to obtain the fermionic UOLEA, it is useful to decompose the interaction matrix $X[\phi]$ into scalar, pseudoscalar, vector, axial-vector and tensor parts. As we restrict our scope to renormalizable UV theories here, we exclude the tensor coupling, and write
\begin{align}
X[\phi] = W_0[\phi] + i W_1[\phi]\gamma^5 + V_{\mu}[\phi]\gamma^{\mu} + A_{\mu}[\phi]\gamma^{\mu}\gamma^5 \ ,
\label{eq:XHdecomp}
\end{align}
where the $W_0,~W_1,~V_\mu,~A_\mu$ coupling matrices are Hermitian.
Obtaining the effective action  for the UV lagrangian above is performed in the standard way, by integrating out the heavy fermion $\Psi$:
\begin{align}
e^{iS_{\rm eff}[\phi]} &= \int \mathcal{D}\overline{\Psi}\mathcal{D}\Psi \, e^{iS_{\rm UV}[\phi,\Psi]} 
\nonumber\\
&\simeq e^{iS_{\rm UV}[\phi, \Psi_c]} \int \mathcal{D}\bar{\eta} \, \mathcal{D}\eta \, e^{i\int d^dx \, \bar{\eta} \left( \slashed{P} - M - X[\phi] \, \right) \eta }
\nonumber \\ 
&= e^{iS_{\rm UV}[\phi, \Psi_c]}  \det \left( \slashed{P} - M - X[\phi]  \right)
= e^{iS_{\rm UV}[\phi, \Psi_c] + \text{Tr}\ln \left( \slashed{P} - M - X[\phi]  \right) } \,  . 
\end{align}
In going from the first to the second line, we have expanded the heavy fields around their classical background values, $\Psi = \Psi_c + \eta$, so that the integration is performed over the quantum fluctuations $\eta$, around the UV action evaluated at this classical solution.
We therefore arrive at the one-loop effective action arising from integrating out heavy fermions:
\begin{align}
S_{\rm eff}^{\text{1-loop}} &= - i \, \text{Tr} \ln \left( \slashed{P} - M - X[\phi]  \, \right),
\end{align} 
where ``Tr'' denotes a trace over both internal indices and over the functional space of the operator $\left( \slashed{P} - M - X[\phi]  \, \right)$. We then evaluate the functional trace by making use of the momentum eigenstate basis, and employing the standard trick of inserting the identity,
\begin{align}
S_{\rm eff}^{\text{1-loop}} &= - i \, \int\dfrac{d^dq}{(2\pi)^d}  \bra{q}  \text{tr} \ln \left( \slashed{P} - M - X[\phi]  \, \right) \ket{q}
\nonumber \\
&= - i \, \int d^dx  \int \dfrac{d^dq}{(2\pi)^d}  \braket{q }{x} \bra{x} \text{tr} \ln \left( \slashed{P} - M - X[\phi]  \, \right) \ket{q}
\nonumber \\
&=-i \int d^dx  \int \dfrac{d^dq}{(2\pi)^d} \text{tr} \ln \left(  \slashed{P} - \slashed{q} - M - X[\phi] \,  \right),
\label{eq:EFT-action-tr}
\end{align}
where now ``tr" denotes a trace over internal indices only. In the last line of (\ref{eq:EFT-action-tr}), we have used $\braket{x}{q}=e^{-iq\cdot x}$ and made a conventional change in the integration variable $q\rightarrow -q$. Further details of these functional manipulations are reviewed in Refs.\cite{Henning:2014wua, Zhang:2016pja}. 

The one-loop effective action of Eq.~(\ref{eq:EFT-action-tr}) must then be expanded in the hard region, where the loop momenta $q^2 \sim M^2$, to obtain the low-energy effective Lagrangian consisting of local operators, as explained, for example, in Refs.~\cite{ Fuentes-Martin:2016uol, Zhang:2016pja}. This method of regions ensures that both heavy-only and mixed heavy-light loops are correctly accounted for in the matching calculation. In the present case, we obtain
\begin{align}
\mathcal{L}^\text{1-loop}_{\rm eff} 
&=- i  \int \left. \dfrac{d^dq}{(2\pi)^d} \text{tr} \ln \left(  \slashed{P} - \slashed{q} - M - X[\phi] \,  \right) \right|_{\rm hard}
\nonumber \\
&=  i \, \text{tr} \sum_{n=1}^{\infty} \dfrac{1}{n} \int \dfrac{d^dq}{(2\pi)^d} \left[ \dfrac{-1}{\slashed{q}+M} \left( - \slashed{P} + W_0[\phi] + i W_1[\phi] \gamma^5 + V_{\mu}[\phi]\gamma^{\mu} + A_{\mu}[\phi]\gamma^{\mu}\gamma^5  \right)  \right]^n \, .
\label{eq:EFT-Lagrangian-tr}
\end{align}
The second equality makes explicit the universal operator structures that appear in the one-loop effective action, and hints at the universality of the corresponding operator coefficients. After expansion and computation of the integrals over the loop momenta, this expression is clearly the fermionic analog of the familiar expression for the bosonic UOLEA of Eq.~\eqref{eq:heavyonlybosonicUOLEA}. We can also see that by virtue of separating $X$ into the sum over the $W_0$, $W_1$, $V_\mu$ and $A_\mu$ components, we can apply our physical intuition for what types of combinations of structures can appear both in the UOLEA itself, and when considering specific models. This will become more clear in the rest of this section, where we discuss the universal structures in more depth, and in Sec.~\ref{sec:examples} when we apply the fermionic UOLEA to several examples.

\subsection{Universal operator structures in the fermionic UOLEA}

In the previous subsection we have described how to obtain a general expression for the fermionic UOLEA. However, as written in Eq.~\eqref{eq:EFT-Lagrangian-tr}, the utility of the UOLEA is not yet apparent. 

It is important to recall that an attractive feature of the bosonic UOLEA is that once the analog of Eq.~\eqref{eq:EFT-Lagrangian-tr} is expanded out to obtain e.g. Eq.~\eqref{eq:heavyonlybosonicUOLEA} (for heavy-only loop contributions to EFT operators up to dimension 6), all necessary structures in the one-loop effective Lagrangian are known and enumerated, and their universal coefficients are calculated once-and-for-all. Having all the possible structures enumerated makes for intuitive application to integrating out particles in specific UV models. Knowing the specific form of the interaction matrix $U$ of Eq.~\eqref{eq:heavyonlybosonicUOLEA} for the UV model being studied allows for dramatic simplification of computation of the one-loop effective action, since not all the bosonic UOLEA operators would contribute to the specific EFT operators of interest. As a trivial example, let us consider a quartic $|\Phi|^2|\phi|^2$ interaction in the UV, such that $U \sim |\phi|^2$. If one is interested in the bosonic UOLEA at dimension 6, it is evident that the term $f_{19}\, U^6$ in Eq.~\ref{eq:heavyonlybosonicUOLEA} is of higher dimension so it will not contribute, and therefore $U^6$ can be discarded without being computed.

Turning to the Fermionic UOLEA, from the form of Eq.~\eqref{eq:EFT-Lagrangian-tr}, we can see that ultimately there will be a proliferation of universal structures in the final one-loop effective Lagrangian, which can be written  compactly as
\begin{align}
    \mathcal{L}_\text{UOLEA}^\text{fermionic} =&\; \sum_{N} f_N\, \Op^{\{P,W_0,W_1,A,V\}}_N \ .
\end{align}
Due to the variety of matrix coupling structures denoted in the superscript set, the fermionic heavy-only UOLEA has a large number of operators in the sum arising from all the (non-vanishing) combinatorial possibilities, in contrast to the bosonic heavy-only UOLEA's 19 operator structures.
An expanded sum of the UOLEA operator classes is presented in Eq.~\eqref{eq:UOLEA_fermionic} and Tables~\ref{tab:W0W1P}-\ref{tab:Axial-only}, where we enumerated all the different classes of possible UOLEA operator structures.

The advantage of separating $X$ into $W_0, W_1, V, A$ is now apparent: all possible universal fermionic UOLEA operators are obtained and their coefficients computed and tabulated once and for all, analogously to the bosonic UOLEA. When inserting a UV model into the fermionic UOLEA, computation of the $W_0, W_1, V, A$ structures then allows for transparent power counting, as well as enabling simple symmetry cross-checks. We now describe this in more detail for each of these (non-vanishing) structures and their combinations listed in Tables~\ref{tab:W0W1P} and \ref{tab:Axial-only}.

\begin{table}[t]
\begin{center}
\scalebox{0.9}{
\begin{tabular}{|c|c|}
\hline
\T\B Operator class & Non-vanishing structures \\
\hline\hline
\T\B $\Op^{(P)}$ & $ P^4,~ P^6$ \\
\hline\hline
\T\B $\Op^{(W_0)}$ & $ W_0, W_0^2, W_0^3, W_0^4, W_0^5, W_0^6$ \\
\hline
\T\B $\Op^{(W_1)}$ & $ W_1^2, W_1^4,  W_1^6$ \\
\hline
\T\B $\Op^{(W_0 W_1)}$ & $W_0W_1^2, W_0^2W_1^2, W_0^3W_1^2, W_0W_1^4, W_0^4W_1^2, W_0^2W_1^4$  \\
\hline
\T\B $\Op^{(PW_0)}$ & $P^2W_0^2, P^2W_0^3, P^4W_0, P^2W_0^4, P^4W_0^2$ \\
\hline
\T\B $\Op^{(P W_1)}$ & $ P^2W_1^2,  P^4W_1, P^2W_1^4, P^4W_1^2$ \\
\hline
\T\B $\Op^{( PW_0W_1 )}$ & $P^2W_0W_1^2, P^4W_0W_1, P^2W_0^2W_1^2$  \\
\hline\hline
\T\B $\Op^{(V)}$ & $ V^2, V^4, V^6$ \\
\hline
\T\B $\Op^{(A)}$ & $ A^2, A^4,  A^6$ \\
\hline
\T\B $\Op^{(V A)}$ & $ VA^3, V^2A^2, V^3A, VA^5, V^2A^4, V^3A^3,  V^4A^2, V^5A$  \\
\hline
\T\B $\Op^{(PV)}$ & $ PV^3, P^2V^2, P^3V, PV^5, P^2V^4, P^3V^3, P^4V^2, P^5V $ \\
\hline
\T\B $\Op^{(P A)}$ & $ PA^3, P^2A^2, P^3A, PA^5, P^2A^4, P^3A^3, P^4A^2, P^5A $ \\
\hline
\T\B \multirow{3}{*}{$\Op^{(P A V)}$} & $PAV^2, PA^2V, P^2AV, PAV^4, PA^2V^3, PA^3V^2, PA^4V,$ \\
\T\B & $P^2AV^3, P^2A^2V^2, P^2A^3V, P^3AV^2, P^3A^2V, P^4AV$ \\
\hline
\end{tabular}
}
\caption{
Non-vanishing operator structures in the fermionic UOLEA that involve covariant derivatives ($P$) plus either scalar and pseudo-scalar structures ($W_0, W_1$), or vector and axial-vector structures ($V, A$).
}
\label{tab:W0W1P}
\end{center}
\end{table}

%================================================%
%               UNIVERSAL STRUCTURES             %
%================================================%
\begin{table}[th!]
\begin{center}
\scalebox{0.9}{
\begin{tabular}{|c|c|}
\hline
\T\B Operator class & Non-vanishing structures \\
\hline\hline
\T\B $\Op^{(V W_0)}$ & $ V^2W_0^2, V^2W_0^3, V^4W_0, V^2W_0^4, V^4W_0^2 $ \\
\hline
\T\B $\Op^{(V W_1)}$ & $  V^2W_1^2, V^4W_1,  V^2W_1^4, V^4W_1^2 $ \\
\hline
\T\B $\Op^{(V W_0 W_1)}$ & $ V^2W_0W_1^2, V^4W_1W_0, V^2W_0^2W_1^2  $ \\
\hline
\T\B \multirow{3}{*}{$\Op^{(P V W_0)}$} & $ PVW_0, PVW_0^2, PVW_0^3, PVW_0^4, PV^3W_0, P^3VW_0, P^2V^2W_0,  $  \\
\T\B & $ P^3VW_0^2, PV^3W_0^2,   P^2V^2W_0^2    $ \\
\hline
\T\B $\Op^{( P V W_1 )}$ & $ PVW_1^2, PVW_1^4, PV^3W_1, P^3VW_1, P^2V^2W_1, PV^3W_1^2, P^3VW_1^2, P^2V^2W_1^2 $  \\
\hline
\T\B $\Op^{(P V W_0 W_1)}$ & $ PVW_0W_1^2, PVW_0^2W_1^2,  P^3VW_0W_1, P^2V^2W_0W_1,  PV^3W_0W_1  $ \\
\hline\hline
\T\B $\Op^{(A W_0)}$ & $ A^2W_0^2, A^2W_0^3, A^4W_0, A^2W_0^4, A^4W_0^2 $ \\
\hline
\T\B $\Op^{(A W_1)}$ & $  A^2W_1^2, A^4W_1,  A^2W_1^4, A^4W_1^2 $ \\
\hline
\T\B $\Op^{(A W_0 W_1)}$ & $ A^2W_0W_1^2, A^4W_1W_0, A^2W_0^2W_1^2  $ \\
\hline
\T\B $\Op^{(P A W_0)}$ & $ PA^3W_0, P^3AW_0, P^2A^2W_0, P^3AW_0^2, PA^3W_0^2, P^2A^2W_0^2 $  \\
\hline
\T\B $\Op^{( P A W_1 )}$ & $ PAW_1, PAW_1^3, PA^3W_1, P^3AW_1, P^2A^2W_1, PA^3W_1^2, P^3AW_1^2, P^2A^2W_1^2 $  \\
\hline
\T\B $\Op^{(P A W_0 W_1)}$ & $ PAW_0W_1, PAW_0^2W_1,  PAW_0W_1^3, PAW_0^3W_1,  P^3AW_0W_1, P^2A^2W_0W_1, PA^3W_0W_1  $ \\
\hline\hline
\T\B $\Op^{(A V W_0)}$ & $ VA^3W_0, V^3AW_0, V^2A^2W_0, V^3AW_0^2, VA^3W_0^2, V^2A^2W_0^2 $ \\
\hline
\T\B $\Op^{(A V W_1)}$ & $ VAW_1, VAW_1^3, VA^3W_1, V^3AW_1, V^2A^2W_1, VA^3W_1^2,  V^3AW_1^2, V^2A^2W_1^2  $ \\
\hline
\T\B $\Op^{(A V W_0 W_1)}$ & $ VAW_0W_1, VAW_0^2W_1, VAW_0W_1^3, VAW_0^3W_1, V^3AW_0W_1, VA^3W_0W_1, V^2A^2W_0W_1 $ \\
\hline
\T\B $\Op^{(P A V W_0)}$ & $ PAV^2W_0, PAV^2W_0^2, PA^2VW_0, PA^2VW_0^2, P^2AVW_0, P^2AVW_0^2  $  \\
\hline
\T\B $\Op^{( P A V W_1 )}$ & $ PAV^2W_1, PAV^2W_1^2, PA^2VW_1, PA^2VW_1^2, P^2AVW_1, P^2AVW_1^2  $  \\
\hline
\T\B $\Op^{(P A V W_0 W_1)}$ & $ PAV^2W_0W_1, PA^2VW_0W_1, P^2AVW_0W_1   $ \\
\hline
\end{tabular}
}
\caption{%\small 
Non-vanishing operator structures in the fermionic UOLEA that involve both (pseudo-)scalar and (axial-)vector couplings.
}
\label{tab:Axial-only}
\end{center}
\end{table}

\subsubsection*{Scalar and pseudo-scalar structures ($W_0$, $W_1$)}

From the Lagrangian as written in Eq.~\eqref{eq:fermionicUVL} and the expansion of $X$ in Eq.~\eqref{eq:XHdecomp}, it is clear that if the heavy fermion that is integrated out has couplings to scalar operators, these will be captured by the $W_0$ matrix structure. Likewise, in the case of couplings to pseudoscalar operators, these will be captured by the $W_1$ matrix. The $W_0$ ($W_1$) matrix is therefore even (odd) under parity, which will allow us to easily intuit what UOLEA operators might be formed and therefore contribute to the final result of Eq.~\eqref{eq:UOLEA_fermionic}. All such structures are listed in Table~\ref{tab:W0W1P}. Indeed, the parity properties of the matrices and their impact on the operator structures is clear. As a scalar structure, $W_0$ can appear in both even and odd powers. In contrast, $W_1$, as a pseudo-scalar structure, must always appear in even powers, or accompanied by $P^4$. That the latter is permitted follows from $\text{tr}(\gamma^\mu\gamma^\nu\gamma^\rho\gamma^\sigma \gamma^5) \neq 0$, so that one can already see that the only EFT operators arising from such a structure will involve pseudo-scalars coupling to $F\tilde{F}$.

\subsubsection*{Vector and axial-vector structures ($V$, $A$)} 

These structures will appear if the UV Lagrangian contains fermionic couplings to vector bosons that do \textit{not} appear in the covariant derivative operator $P$. This would occur, for example, if the heavy fermion current is coupled to a light gauge boson such as the $Z_\mu$ of the SM (in this case the low-energy effective theory with $Z_\mu$ not in a covariant derivative would not be the SMEFT), or an $A^\prime_\mu$ associated with a broken $U(1)^\prime$ whose mass was sufficiently small compared with that of the fermion being integrated out. These results are particularly useful if one is interested in matching to low-energy EFTs containing massive vector bosons. In this case, it should be noted that the covariant derivative operator $P$ only contains the gauge fields associated with the remaining unbroken symmetries.

Even if the gauge boson content of the low-energy theory is purely that of the SM, these structures must be included in a complete fermionic UOLEA if one wishes to apply it to matching with general EFTs. 
We will see examples in Sec.~\ref{sec:examples} where the $V$ and $A$ structures appear.

As in the case above of $W_0$, $W_1$ operator structures, the power counting and enumeration of non-vanishing $V$ and $A$ combinations is straightforwardly obtained from symmetry arguments. All structures that contribute to EFT operators up to dimension 6 are listed in Table~\ref{tab:W0W1P}. We can see that by virtue of the symmetry properties of both $V$ and $A$, they must always appear in the combinations $P^k V^l A^m$ with $k+l+m$ even.

\subsubsection*{General case ($W_0$, $W_1$, $V$, $A$ all present)}

The above discussion can be extended further to the situation when all possible structures in Eq.~\eqref{eq:fermionicUVL} are present. In this most general case, one gets a proliferation of possible combinations and operator classes, all of which are listed in Table~\ref{tab:Axial-only}.

Once again, the power counting is straightforward, and follows from trace identities of gamma matrices. As before, scalar structures $W_0$ can appear without restriction, while pseudo-scalar structures $W_1$ can only appear in combination with operators such that the overall number of $\gamma^5$ matrices is even, or in combination with four $\gamma^\mu$ matrices.

\subsection{Computing UOLEA operators with covariant diagrams}
\label{sec:computationUOLEA}

To evaluate the expansion in \eqref{eq:EFT-Lagrangian-tr}, we use the covariant diagrams technique of Ref.\cite{Zhang:2016pja} to keep track of the expansion and directly compute the Wilson coefficient for each EFT operator. Each term in the CDE expansion \eqref{eq:EFT-Lagrangian-tr} can be represented by a covariant diagram, which can be written down directly by a systematic set of rules. We then straightforwardly obtain the prefactor coefficient and the one-loop master integral associated with the diagram we are considering. The details of the covariant diagram technique are described in Ref.\cite{Zhang:2016pja}. Here we summarise the essential ingredients relevant for the present case of heavy fermion loops. 
\begin{itemize}
\item \textbf{Fermion propagator:}\\
Each fermion propagator can be decomposed into two terms,
\begin{align}
\dfrac{-1}{\slashed{q}+M} = \dfrac{M}{q^2-M^2} + \dfrac{-q_{\mu}\gamma^{\mu}}{q^2-M^2}, 
\label{eq:propagator-Fermionic}
\end{align}
where the first term is the heavy bosonic propagator multiplied by the mass. The second term involves the loop momentum $q_{\mu}$ in the numerator, which will contribute to the loop integral. The loop integrals have the general form
\begin{align}
\int \dfrac{d^dq}{(2\pi)^d} \dfrac{q^{\mu_1}\cdots q^{\mu_{2n_c}}}{(q^2-M_i^2)^{n_i}(q^2-M_j^2)^{n_j}\cdots (q^{2})^{n_L}}
= g^{\mu_1 \cdots \mu_{2n_c}} \mathcal{I}[q^{2n_c}]^{n_i n_j \cdots n_L}_{i j \cdots 0},
\label{eq:master-Integral}
\end{align}
where $g^{\mu_1 \cdots \mu_{2n_c}}$ is the completely symmetric tensor, e.g. $g^{\mu\nu\rho\sigma} = g^{\mu\nu}g^{\rho\sigma}+g^{\mu\rho}g^{\nu\sigma}+g^{\mu\sigma}g^{\nu\rho}$, and $\mathcal{I}$ are master integrals, a useful set of which can be found in Ref.~\cite{Zhang:2016pja}. The symmetric tensor in \eqref{eq:master-Integral} will contract the Lorentz indices of Dirac matrices in the fermionic propagator, then we must sum over all possibilities of the contractions. In the covariant diagram, we shall use dotted lines to indicate the contractions among the fermionic part of the propagator in Eq.~\eqref{eq:propagator-Fermionic}, following the conventions of Ref.~\cite{Zhang:2016pja}. 
\item \textbf{Vertex insertions:} From Eq.~\eqref{eq:EFT-Lagrangian-tr}, all vertex insertions, $\gamma^\mu P_\mu, W_0, i\gamma^5 W_1, \gamma^{\mu}V_{\mu}$ and $ \gamma^{\mu}\gamma^5A_{\mu}$ are independent of the loop momentum $q_{\mu}$ and thus do not change the loop integrals. We note that the vertex insertions will not be contracted with each other or with the propagators. 
\item \textbf{Dirac trace evaluations:} By construction, $P_{\mu}, W_0, W_1, V_{\mu}$ and $A_{\mu}$ do not involve additional Dirac matrices. Therefore, after reading off the value of each covariant diagram, the $trace$ over Dirac matrices is factorized out and evaluated once-and-for-all. The trace in the final results, still denoted by ``tr'' is over the remaining internal indices, e.g.\ $SU(2)$ and color indices. 
\item \textbf{Renormalisation and divergences:} For the one-loop divergent integrals, we use dimensional regularisation and the $\overline{\text{MS}}$-scheme for renormalisation. The important point in the case with divergent integrals is that the $trace$ over all Dirac matrices have to be evaluated in $D=4-\epsilon$ dimensions, and the $\epsilon$-term resulting from the contractions of the metric tensor, $g_{\mu\nu}g^{\mu\nu} = D$, must be kept in the computations. This term can hit the $1/\epsilon$ pole resulting from a divergent integral and yield a finite contribution to the Wilson coefficient. 
It is well-known that in $D=4-\epsilon$ dimensions, the relations $\{ \gamma^\mu, \gamma^5\} = 0$ and $\text{tr}(\gamma^\mu \gamma^\nu \gamma^\rho \gamma^\sigma \gamma^5) \neq 0$ cannot be satisfied simultaneously~\cite{Chanowitz:1979zu, Jegerlehner:2000dz}. 
In our calculations, we use the Breitenlohner-Maison-'t Hooft-Veltman (BMHV) scheme~\cite{tHooft:1972tcz,Breitenlohner:1977hr}.
\item\textbf{Covariant derivatives in commutators:} By expanding the one-loop effective action in Eq.\eqref{eq:EFT-Lagrangian-tr}, we will obtain operator structures that carry ``open" covariant derivatives, $P_{\mu}$. We emphasise that the $P_{\mu}$ in the CDE expansion is a functional operator, i.e.\ $P_{\mu}$ will act on everything to the right. To construct an effective operator we need a ``closed" covariant derivative where $P_{\mu}$ will only act on its immediate nearest neighbour in the operator; we thus need to organise the final results such that $P_{\mu}$'s only appear in commutators (see e.g.\ Refs.~\cite{Henning:2014wua, Zhang:2016pja, Cohen:2019btp}). To be concrete, let us consider for example a functional operator $P_{\mu}W_0$ acting on a generic functional $\phi$:
\begin{align}
   P_{\mu}W_0 ~\phi = \left(P_{\mu}W_0\right)_{\rm local}\phi + W_0\left(P_{\mu}\phi\right), 
\end{align}
we then combine all operator structures with $P_{\mu}$ into commutators,
\begin{align}
    \left(P_{\mu}W_0\right)_{\rm local}\phi = \left(P_{\mu}W_0 - W_0P_{\mu} \right)\phi = \left[P_{\mu}, W_0 \right]\phi \, .
\end{align}
In practice, we first write down a basis set of independent operators where $P_{\mu}$'s only appear in the commutators, and then expand the commutators and match the results from the functional trace expansion to solve the system of equations and determine the coefficient of the elements in the ``commutator" basis that we chose. We note that the operator structures with adjacent covariant derivatives, $\text{tr}\left( \cdots P^2 \cdots \right)$, can be dropped to simplify this computation, since the non-$P^2$ terms are sufficient for reconstructing the universal operators written in the commutator basis when matching to the expanded form (see Ref.~\cite{Zhang:2016pja} for details). 

\item \textbf{Hermiticity of the operator structures:} Since covariant diagrams that are mirror images of each other are related by hermitian conjugation, only one in each such pair needs to be computed. We will also use the hermiticity of the Lagrangian to identify the number of irreducible operator structures. In particular, when the vector and axial-vector structures are included in the matrix $X[\phi]$, the hermitian conjugate relations can drastically reduce the number of operator structures we need to evaluate.
\end{itemize}
Let us consider a simple example to illustrate concretely some of these points, taking a coupling matrix $X[\phi]$ that only contains pseudo-scalar structures. We would then compute the universal coefficient of the operator structure $P^2W_1^2$ as follows:
\begin{align}
\Op^{P^2W_1^2} =& 
\vcenter{
 \hbox{\includegraphics[scale=0.75]{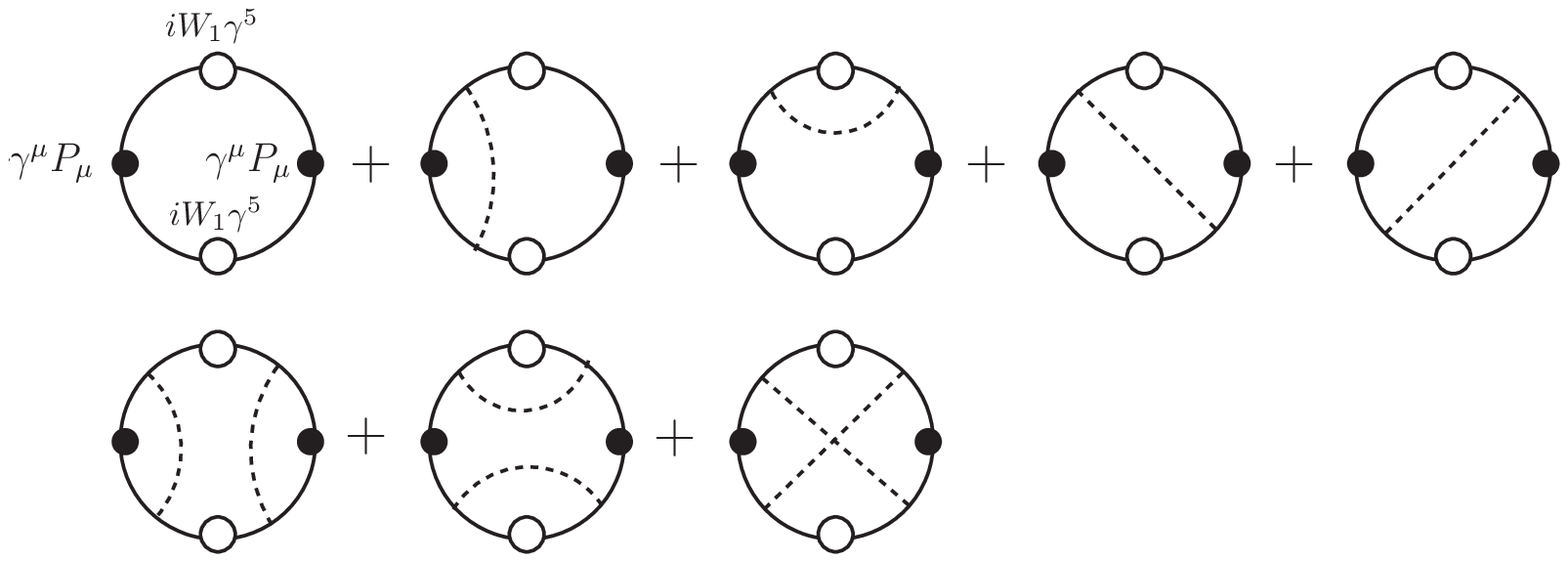}}	} 
\nonumber \\[4pt]
=& \, \frac{i}{2} m^4 \mathcal{I}_i^4 \, \text{tr}\left( \slashed{P} iW_1 \gamma^5 \slashed{P} iW_1\gamma^5 \right)
\nonumber \\
&+ im^2\mathcal{I}[q^2]_i^4 
\left[ 
\text{tr}\left( \gamma^{\mu}\slashed{P}\gamma_{\mu} iW_1\gamma^5 \slashed{P} iW_1\gamma^5 \right) 
+ \text{tr}\left( \slashed{P}\gamma^{\mu} iW_1\gamma^5 \gamma_{\mu} \slashed{P} iW_1\gamma^5 \right) \right.
\nonumber\\
&\qquad\qquad\qquad \left. + \dfrac{1}{2}\text{tr}\left( \slashed{P} \gamma^{\mu} iW_1\gamma^5 \slashed{P} \gamma_{\mu} iW_1\gamma^5 \right)
+ \dfrac{1}{2}\text{tr}\left( \slashed{P} iW_1\gamma^5 \gamma^{\mu} \slashed{P} iW_1\gamma^5 \gamma_{\mu} \right)
\right]
\nonumber\\
&+ i\mathcal{I}[q^4]_i^4 
\left[ \dfrac{1}{2}\text{tr}\left( \gamma^{\mu}\slashed{P}\gamma_{\mu} iW_1\gamma^5 \gamma^{\nu}\slashed{P}\gamma_{\nu} iW_1\gamma^5 \right) 
+ \dfrac{1}{2}\text{tr}\left( \slashed{P} \gamma^{\mu} iW_1\gamma^5 \gamma_{\mu} \slashed{P} \gamma^{\nu} iW_1\gamma^5 \gamma_{\nu} \right)
\right.
\nonumber\\
&\qquad\qquad~ \left. + \dfrac{1}{2}\text{tr}\left( \slashed{P} \gamma^{\mu} iW_1\gamma^5 \gamma^{\nu} \slashed{P} \gamma_{\mu} iW_1\gamma^5 \gamma_{\nu} \right)
\right]
\nonumber\\
&= i\left( 2m^4\mathcal{I}_i^4 - 16m^2\mathcal{I}[q^2]_i^4 + \left[ 48-4\epsilon \right]\mathcal{I}[q^4]_i^4 \right) \text{tr}\left( P_{\mu}W_1P_{\mu}W_1 \right),
\label{example:p2w12}
\end{align}
where the loop integral $\mathcal{I}[q^4]_i^4$ is divergent and thus we evaluated the Dirac trace in $D=4-\epsilon$ dimensions and kept the $\mathcal{O(\epsilon})$ terms. Note that we have omitted diagrams where the two $\slashed{P}$ insertions are adjacent, because they lead to terms proportional to $\text{tr}(\dots P^2\dots)$, which provide redundant information for constructing independent operators as discussed above. Finally, we re-write the operator structures in Eq.~\eqref{example:p2w12} in terms of commutators, using
\begin{align}
    2f_N^{(P^2W_1^2)} \, \text{tr} \left( P_{\mu}W_1P_{\mu}W_1 \right)
    \supset f_N^{(P^2W_1^2)} \, \text{tr} \left( \left[P_{\mu},W_1\right] \left[P_{\mu},W_1\right] \right) \, ,
    \label{example:p2w12commutator}
\end{align}
 and therefore obtain the final result
\begin{align}
    \mathcal{L}_{\rm EFT}^{\rm 1-loop}[\phi] &\supset i\left( m^4\mathcal{I}_i^4 - 8m^2\mathcal{I}[q^2]_i^4 + \left[ 24-2\epsilon \right]\mathcal{I}[q^4]_i^4 \right) \text{tr} \left( \left[P_{\mu},W_1\right] \left[P_{\mu},W_1\right] \right)
    \nonumber\\
    &\supset \dfrac{i^2}{(4\pi)^2}\left( -\log\dfrac{m^2}{\mu^2} + \dfrac{2}{3} \right) \text{tr} \left( \left[P_{\mu},W_1\right] \left[P_{\mu},W_1\right] \right) \, ,
\end{align}
 making use of the master integrals listed in Ref.~\cite{Zhang:2016pja}.

%--Pure-Gauge Operators------------------------------%
\begin{table}[htpb!]
\begin{center}
\scalebox{0.9}{
\begin{tabular}{|c|c|}
\hline
\multicolumn{2}{|c|}{\T\B $\Op^{(P)}$ terms  }  \\
\hline 
%\hline
\T\B $  - \frac{1}{2} \mathcal{I}_i^4 m_i^4 + 4 m_i^2 \mathcal{I}[q^2]_i^4 + (5 \epsilon  - 8) \mathcal{I}[q^4]_i^4 $ & $ [P_{\mu},P_{\nu}] [P_{\mu},P_{\nu}] $ \\ 
\hline
%-----------------Single Table of Operators------------------%
\T\B $ 24 m_i^2 \mathcal{I}[q^4]_i^6 - 2 m_i^4 \mathcal{I}[q^2]_i^6 - 64 \mathcal{I}[q^6]_i^6 $ & $ [P_{\mu},[P_{\mu},P_{\nu}]] [P_{\rho},[P_{\rho},P_{\nu}]] $ \\ 
\T\B $  - \frac{2}{3} \mathcal{I}_i^6 m_i^6 + 4 m_i^4 \mathcal{I}[q^2]_i^6 - \frac{128}{3} \mathcal{I}[q^6]_i^6 $ & $ [P_{\mu},P_{\nu}] [P_{\nu},P_{\rho}] [P_{\rho},P_{\mu}] $ \\ 
 \hline 
\end{tabular}
}
\caption{Pure gauge operator structures in the fermionic UOLEA.}
\label{tab:pureP}
\end{center}
\end{table}
%-----------------Single Table of Operators------------------%
\begin{table}[ht!]
\begin{center}
\scalebox{0.9}{
\begin{tabular}{|c|c|}
\hline
\multicolumn{2}{|c|}{\T\B $\Op^{(W_0)}$ terms }  \\
\hline 
\T\B $ 4m_i \mathcal{I}_i $ & $ W_0 $ \\ 
\hline 
\T\B $ 2 \mathcal{I}_i^2 m_i^2 + (8 - 2 \epsilon ) \mathcal{I}[q^2]_i^2 $ & $ W_0^2 $ \\ 
 \hline 
\T\B $ \frac{4}{3} \mathcal{I}_i^3 m_i^3 + ( \, 16 m_i - 4 \epsilon  m_i  \, ) \mathcal{I}[q^2]_i^3 $ & $ W_0^3 $ \\ 
 \hline 
\T\B $ \mathcal{I}_i^4 m_i^4 + 24 m_i^2 \mathcal{I}[q^2]_i^4 + (24 - 10 \epsilon ) \mathcal{I}[q^4]_i^4 $ & $ W_0^4 $ \\ 
 \hline 
\T\B $ \frac{4}{5} \mathcal{I}_i^5 m_i^5 + 96 m_i \mathcal{I}[q^4]_i^5 + 32 m_i^3 \mathcal{I}[q^2]_i^5 $ & $ W_0^5 $ \\ 
 \hline 
\T\B $ \frac{2}{3} \mathcal{I}_i^6 m_i^6 + 240 m_i^2 \mathcal{I}[q^4]_i^6 + 40 m_i^4 \mathcal{I}[q^2]_i^6 + 128 \mathcal{I}[q^6]_i^6 $ & $ W_0^6 $ \\ 
\hline 
\end{tabular}
}
\end{center}
%--------------------Mixed P-W0----------------------%
\begin{center}
\scalebox{0.9}{
\begin{tabular}{|c|c|}
\hline
\multicolumn{2}{|c|}{\T\B $\Op^{(PW_0)}$ terms }  \\
\hline 
%\hline
\T\B $ \mathcal{I}_i^4 m_i^4 + (24 - 10 \epsilon ) \mathcal{I}[q^4]_i^4 $ & $ [P_{\mu},W_0] [P_{\mu},W_0] $ \\ 
 \hline
%-----------------Single Table of Operators------------------%
\T\B $ 4 \mathcal{I}_i^5 m_i^5 + 192 m_i \mathcal{I}[q^4]_i^5 + 16 m_i^3 \mathcal{I}[q^2]_i^5 $ & $ W_0 [P_{\mu},W_0] [P_{\mu},W_0] $ \\ 
% \hline 
%-----------------Single Table of Operators------------------%
\hline
\T\B $  - 2 \mathcal{I}_i^5 m_i^5 - 16 m_i \mathcal{I}[q^4]_i^5 + 16 m_i^3 \mathcal{I}[q^2]_i^5 $ & $ W_0 [P_{\mu},P_{\nu}] [P_{\mu},P_{\nu}] $ \\ 
\hline 
%------------------Single Table of Operators----------------%
\T\B $ 4 \mathcal{I}_i^6 m_i^6 + 432 m_i^2 \mathcal{I}[q^4]_i^6 + 36 m_i^4 \mathcal{I}[q^2]_i^6 + 192 \mathcal{I}[q^6]_i^6 $ & $ W_0 [P_{\mu},W_0] W_0 [P_{\mu},W_0] $ \\ 
\T\B $ 6 \mathcal{I}_i^6 m_i^6 + 576 m_i^2 \mathcal{I}[q^4]_i^6 + 60 m_i^4 \mathcal{I}[q^2]_i^6 + 576 \mathcal{I}[q^6]_i^6 $ & $ W_0^2 [P_{\mu},W_0] [P_{\mu},W_0] $ \\ 
 \hline 
%-----------------Single Table of Operators------------------%
\T\B $ 2 \mathcal{I}_i^6 m_i^6 - 16 m_i^2 \mathcal{I}[q^4]_i^6 - 16 m_i^4 \mathcal{I}[q^2]_i^6 $ & $ [P_{\mu},W_0] [P_{\nu},W_0] [P_{\mu},P_{\nu}] $ \\ 
\T\B $  - 5 \mathcal{I}_i^6 m_i^6 + 72 m_i^2 \mathcal{I}[q^4]_i^6 + 36 m_i^4 \mathcal{I}[q^2]_i^6 - 64 \mathcal{I}[q^6]_i^6 $ & $ W_0^2 [P_{\mu},P_{\nu}] [P_{\mu},P_{\nu}] $ \\ 
\T\B $  - 2 \mathcal{I}_i^6 m_i^6 - 8 m_i^2 \mathcal{I}[q^4]_i^6 + 18 m_i^4 \mathcal{I}[q^2]_i^6 + 96 \mathcal{I}[q^6]_i^6 $ & $ \left(W_0 [P_{\mu},W_0]-[P_{\mu},W_0] W_0\right) [P_{\nu},[P_{\mu},P_{\nu}]] $ \\ 
\T\B $ 8 m_i^2 \mathcal{I}[q^4]_i^6 + 2 m_i^4 \mathcal{I}[q^2]_i^6 + 96 \mathcal{I}[q^6]_i^6 $ & $ [P_{\mu},[P_{\mu},W_0]] [P_{\nu},[P_{\nu},W_0]] $ \\ 
 \hline 
\end{tabular}
}
\caption{Operator structures in the degenerate fermionic UOLEA involving the scalar coupling $W_0$.}
\label{tab:mixedPW0}
\end{center}
\end{table}
\clearpage

%--Pure W1 operators----------------------------------------------%

%-----------------Single Table of Operators------------------%
\begin{table}[t]%[ht!]
\begin{center}
\scalebox{0.9}{
\begin{tabular}{|c|c|}
\hline
\multicolumn{2}{|c|}{\T\B 
$\Op^{(W_1)}$ terms }  \\
\hline 
%\hline
\T\B $ 2 (\epsilon  + 4) \mathcal{I}[q^2]_i^2 - 2 \mathcal{I}_i^2 m_i^2 $ & $ W_1^2 $ \\ 
 \hline 
\T\B $ \mathcal{I}_i^4 m_i^4 - 8 m_i^2 \mathcal{I}[q^2]_i^4 + 2 (11 \epsilon  + 12) \mathcal{I}[q^4]_i^4 $ & $ W_1^4 $ \\ 
 \hline 
\T\B $  - \frac{2}{3} \mathcal{I}_i^6 m_i^6 - 48 m_i^2 \mathcal{I}[q^4]_i^6 + 8 m_i^4 \mathcal{I}[q^2]_i^6 + 128 \mathcal{I}[q^6]_i^6 $ & $ W_1^6 $ \\ 
 \hline 
\end{tabular}
}
\end{center}
%--Mixed P-W1 terms---------------------------------------------%
\begin{center}
\scalebox{0.9}{
\begin{tabular}{|c|c|}
\hline
\multicolumn{2}{|c|}{\T\B $\Op^{(PW_1)}$ terms }  \\
\hline 
\T\B $ \mathcal{I}_i^4 m_i^4 - 8 m_i^2 \mathcal{I}[q^2]_i^4 - 2 (\epsilon  - 12) \mathcal{I}[q^4]_i^4 $ & $ [P_{\mu},W_1] [P_{\mu},W_1] $ \\ 
\hline 
%=======Merge 2 table======%
\T\B $ 24 m_i \mathcal{I}[q^4]_i^5  - 8 m_i^3 \mathcal{I}[q^2]_i^5  + \mathcal{I}_i^5 m_i^5 $ & $ \epsilon_{\mu \nu \rho \sigma} \, W_1 [P_{\mu},P_{\nu}] [P_{\rho},P_{\sigma}] $ \\ 
\hline 
%===============Merge table=============%
\T\B $  - 48 m_i^2 \mathcal{I}[q^4]_i^6 + 4 m_i^4 \mathcal{I}[q^2]_i^6 + 192 \mathcal{I}[q^6]_i^6 $ & $ W_1 [P_{\mu},W_1] W_1 [P_{\mu},W_1] $ \\ 
\T\B $  - 2 \mathcal{I}_i^6 m_i^6 - 192 m_i^2 \mathcal{I}[q^4]_i^6 + 28 m_i^4 \mathcal{I}[q^2]_i^6 + 576 \mathcal{I}[q^6]_i^6 $ & $ W_1^2 [P_{\mu},W_1] [P_{\mu},W_1] $ \\ 
\hline 
 %=============Merge tables==========%
\T\B $  - 2 \mathcal{I}_i^6 m_i^6 - 48 m_i^2 \mathcal{I}[q^4]_i^6 + 16 m_i^4 \mathcal{I}[q^2]_i^6 $ & $ [P_{\mu},W_1] [P_{\nu},W_1] [P_{\mu},P_{\nu}] $ \\ 
\T\B $ \mathcal{I}_i^6 m_i^6 + 56 m_i^2 \mathcal{I}[q^4]_i^6 - 12 m_i^4 \mathcal{I}[q^2]_i^6 - 64 \mathcal{I}[q^6]_i^6 $ & $ W_1^2 [P_{\mu},P_{\nu}] [P_{\mu},P_{\nu}] $ \\ 
\T\B $  - 24 m_i^2 \mathcal{I}[q^4]_i^6 + 2 m_i^4 \mathcal{I}[q^2]_i^6 + 96 \mathcal{I}[q^6]_i^6 $ & $ [P_{\mu},[P_{\mu},W_1]] [P_{\nu},[P_{\nu},W_1]] $ \\ 
\T\B $  - 24 m_i^2 \mathcal{I}[q^4]_i^6 + 2 m_i^4 \mathcal{I}[q^2]_i^6 + 96 \mathcal{I}[q^6]_i^6 $ & $ \left(W_1 [P_{\mu},W_1]-[P_{\mu},W_1] W_1\right) [P_{\nu},[P_{\mu},P_{\nu}]] $ \\ 
 \hline 
\end{tabular}
}
\caption{Operator structures in the degenerate fermionic UOLEA involving the pseudoscalar coupling $W_1$.}
\label{tab:mixedPW1}
\end{center}
\end{table}

%=======================================================
%--Mixed W0-W1 terms

\begin{table}[ht!]

\scalebox{0.605}{
\begin{tabular}{|c|c|}
    \hline
\multicolumn{2}{|c|}{\T\B $\Op^{(W_0W_1)}$ terms } 
    \\ \hline

\T\B $ 4 (3 \epsilon  + 4) m_i \mathcal{I}[q^2]_i^3 - 4 \mathcal{I}_i^3 m_i^3 $ & $W_0 W_1^2$
\\
\hline
  
\T\B $ 8 (\epsilon  + 12) \mathcal{I}[q^4]_i^4 - 4 \mathcal{I}_i^4 m_i^4 $ 
    & $W_0^2W_1^2$ 
\\
 
\T\B $  - 2 \mathcal{I}_i^4 m_i^4 + 16 m_i^2 \mathcal{I}[q^2]_i^4 + 4 (5 \epsilon  - 12) \mathcal{I}[q^4]_i^4 $ & $ W_0 W_1 W_0 W_1 $ 
\\
 \hline
 \T\B $  - 4 \mathcal{I}_i^5 m_i^5 + 288 m_i \mathcal{I}[q^4]_i^5 - 32 m_i^3 \mathcal{I}[q^2]_i^5 $ 
& $W_0^3 W_1^2$
\\
\T\B $  - 4 \mathcal{I}_i^5 m_i^5 - 96 m_i \mathcal{I}[q^4]_i^5 + 32 m_i^3 \mathcal{I}[q^2]_i^5 $ 
&  $W_0^2 W_1 W_0 W_1$
\\
\hline
\T\B $ 4 \mathcal{I}_i^5 m_i^5 + 96 m_i \mathcal{I}[q^4]_i^5 - 32 m_i^3 \mathcal{I}[q^2]_i^5 $ & $ W_0 W_1^4 $
\\
\hline
\T\B $  - 4 \mathcal{I}_i^6 m_i^6 + 96 m_i^2 \mathcal{I}[q^4]_i^6 + 16 m_i^4 \mathcal{I}[q^2]_i^6 - 768 \mathcal{I}[q^6]_i^6 $ 
&  $W_0^3 W_1 W_0 W_1$ 
\\ 
\T\B $  - 2 \mathcal{I}_i^6 m_i^6 - 144 m_i^2 \mathcal{I}[q^4]_i^6 + 24 m_i^4 \mathcal{I}[q^2]_i^6 + 384 \mathcal{I}[q^6]_i^6 $ 
    & $ W_0^2 W_1 W_0^2W_1 $
\\
\T\B $  - 4 \mathcal{I}_i^6 m_i^6 + 480 m_i^2 \mathcal{I}[q^4]_i^6 - 80 m_i^4 \mathcal{I}[q^2]_i^6 + 768 \mathcal{I}[q^6]_i^6 $ & $ W_0^4 W_1^2 $
\\
\hline
\T\B $ 4 \mathcal{I}_i^6 m_i^6 + 288 m_i^2 \mathcal{I}[q^4]_i^6 - 48 m_i^4 \mathcal{I}[q^2]_i^6 - 768 \mathcal{I}[q^6]_i^6 $ 
& $W_0 W_1 W_0 W_1^3$
\\
\T\B $ 2 \mathcal{I}_i^6 m_i^6 - 48 m_i^2 \mathcal{I}[q^4]_i^6 - 8 m_i^4 \mathcal{I}[q^2]_i^6 + 384 \mathcal{I}[q^6]_i^6 $ 
&  $W_0 W_1^2 W_0 W_1^2$
\\ 
\T\B $ 4 \mathcal{I}_i^6 m_i^6 - 96 m_i^2 \mathcal{I}[q^4]_i^6 - 16 m_i^4 \mathcal{I}[q^2]_i^6 + 768 \mathcal{I}[q^6]_i^6 $ 
& $ W_0^2 W_1^4 $
\\ 
\hline 
\end{tabular}
}
\scalebox{0.605}{
\begin{tabular}{|c|c|}
\hline
\multicolumn{2}{|c|}{\T\B $\Op^{(PW_0W_1)}$ terms }  \\
\hline 
\T\B $ 48 m_i \mathcal{I}[q^4]_i^5 - 8 m_i^3 \mathcal{I}[q^2]_i^5 $ & $ W_1 [P_{\mu},W_0] [P_{\mu},W_1] + \text{h.c.} $ \\ 
\T\B $ 4 \mathcal{I}_i^5 m_i^5 + 96 m_i \mathcal{I}[q^4]_i^5 - 32 m_i^3 \mathcal{I}[q^2]_i^5 $ & $ W_0 [P_{\mu},W_1] [P_{\mu},W_1] $ \\ 
\hline 
%=============Merge tables==========%
\T\B $ 24 m_i^2 \mathcal{I}[q^4]_i^6  - 8 m_i^4 \mathcal{I}[q^2]_i^6  + \mathcal{I}_i^6 m_i^6 $ & $ \epsilon_{\mu \nu \rho \sigma} \, W_0 W_1 [P_{\mu},P_{\nu}] [P_{\rho},P_{\sigma}] + \text{h.c.} $ \\ 
\T\B $ 24 m_i^2 \mathcal{I}[q^4]_i^6  - 8 m_i^4 \mathcal{I}[q^2]_i^6  + \mathcal{I}_i^6 m_i^6 $ & $\epsilon_{\mu \nu \rho \sigma} \, W_0 [P_{\mu},P_{\nu}] W_1 [P_{\rho},P_{\sigma}] $ \\ 
%========================%
% Big Merge
%========================%
\hline 
\T\B $ 2 \mathcal{I}_i^6 m_i^6 + 192 m_i^2 \mathcal{I}[q^4]_i^6 - 28 m_i^4 \mathcal{I}[q^2]_i^6 - 576 \mathcal{I}[q^6]_i^6 $ & $ W_1 W_0 [P_{\mu},W_1] [P_{\mu},W_0] + \text{h.c.} $ \\ 
\T\B $ 48 m_i^2 \mathcal{I}[q^4]_i^6 - 4 m_i^4 \mathcal{I}[q^2]_i^6 - 192 \mathcal{I}[q^6]_i^6 $ & $ W_1 [P_{\mu},W_0] W_1 [P_{\mu},W_0] $ \\ 
\T\B $ 4 \mathcal{I}_i^6 m_i^6 + 144 m_i^2 \mathcal{I}[q^4]_i^6 - 36 m_i^4 \mathcal{I}[q^2]_i^6 - 192 \mathcal{I}[q^6]_i^6 $ & $ W_0 [P_{\mu},W_1] W_0 [P_{\mu},W_1] $ \\ 
\T\B $ 96 m_i^2 \mathcal{I}[q^4]_i^6 - 24 m_i^4 \mathcal{I}[q^2]_i^6 + 384 \mathcal{I}[q^6]_i^6 $ & $ W_1 [P_{\mu},W_1] W_0 [P_{\mu},W_0] + \text{h.c.} $ \\ 
\T\B $ 6 \mathcal{I}_i^6 m_i^6 - 36 m_i^4 \mathcal{I}[q^2]_i^6 + 576 \mathcal{I}[q^6]_i^6 $ & $ W_0^2 [P_{\mu},W_1] [P_{\mu},W_1] $ \\ 
\T\B $  - 2 \mathcal{I}_i^6 m_i^6 - 4 m_i^4 \mathcal{I}[q^2]_i^6 + 576 \mathcal{I}[q^6]_i^6 $ & $ W_0 W_1 [P_{\mu},W_1] [P_{\mu},W_0] + \text{h.c.} $ \\ 
\T\B $  - 2 \mathcal{I}_i^6 m_i^6 - 4 m_i^4 \mathcal{I}[q^2]_i^6 + 576 \mathcal{I}[q^6]_i^6 $ & $ W_1^2 [P_{\mu},W_0] [P_{\mu},W_0] $ \\ 
 \hline 
\end{tabular}
}
\caption{Operator structures in the degenerate fermionic UOLEA involving both the scalar coupling $W_0$ and the pseudoscalar coupling $W_1$.
}
\label{tab:mixedPW0W1}
\end{table}

%===================================%
% A-V tables
%===================================%
\begin{table}[ht!]
\begin{minipage}[t]{0.45\linewidth}\centering
\scalebox{0.76}{
\begin{tabular}{|c|c|}
\hline
\multicolumn{2}{|c|}{\T\B $\Op^{(P^2 A^2 W_1)}$ terms  }  \\
\hline 
\T\B $ 4m_i^5\mathcal{I}_i^5 - 16m_i^3\mathcal{I}[q^2]_i^5 $ & $ \epsilon^{\mu\nu\rho\sigma} P_{\mu} A_{\nu} P_{\rho} A_{\sigma} W_1 + \text{h.c.} $ 
    \\ 
\T\B $ -4m_i^5\mathcal{I}_i^5 + 16m_i^3\mathcal{I}[q^2]_i^5 $ & $ \epsilon^{\mu\nu\rho\sigma} P_{\mu} P_{\nu} A_{\rho} A_{\sigma} W_1 + \text{h.c.} $
    \\ 
\T\B $ 4m_i^5\mathcal{I}_i^5 - 96m_i\mathcal{I}[q^4]_i^5 $ & $\epsilon^{\mu\nu\rho\sigma} P_{\mu} W_1 P_{\nu} A_{\rho} A_{\sigma} $
    \\ 
\T\B $ 4m_i^5\mathcal{I}_i^5 - 32m_i^3\mathcal{I}[q^2]_i^5 + 96m_i\mathcal{I}[q^4]_i^5 $ & $\epsilon^{\mu\nu\rho\sigma} P_{\mu} P_{\nu} A_{\rho} W_1 A_{\sigma} $
    \\ \hline
\end{tabular}
}
\end{minipage}
\hfill
%====================================%
\begin{minipage}[t]{0.48\linewidth}\centering
\scalebox{0.76}{
\begin{tabular}{|c|c|}
\hline
\multicolumn{2}{|c|}{\T\B $\Op^{(P^2 V^2 W_1)}$ terms  }  \\
\hline 
\T\B $ -4m_i^5\mathcal{I}_i^5 + 32m_i^3\mathcal{I}[q^2]_i^5 - 96m_i\mathcal{I}[q^4]_i^5 $ & $ \epsilon^{\mu\nu\rho\sigma} P_{\mu} W_1 P_{\nu} V_{\rho} V_{\sigma} $ 
    \\ 
\T\B $ -4m_i^5\mathcal{I}_i^5 + 32m_i^3\mathcal{I}[q^2]_i^5 - 96m_i\mathcal{I}[q^4]_i^5 $ & $ \epsilon^{\mu\nu\rho\sigma} P_{\mu} P_{\nu} V_{\rho} W_1 V_{\sigma} $
    \\ 
\T\B $ 4m_i^5\mathcal{I}_i^5 - 32m_i^3\mathcal{I}[q^2]_i^5 + 96m_i\mathcal{I}[q^4]_i^5 $ & $\epsilon^{\mu\nu\rho\sigma} P_{\mu} P_{\nu} V_{\rho} V_{\sigma} W_1 + \text{h.c.} $
    \\ 
\T\B $ 4m_i^5\mathcal{I}_i^5 - 32m_i^3\mathcal{I}[q^2]_i^5 + 96m_i\mathcal{I}[q^4]_i^5 $ & $\epsilon^{\mu\nu\rho\sigma} P_{\mu} V_{\nu} P_{\rho} V_{\sigma} W_1 + \text{h.c.} $
    \\ \hline
\end{tabular}
}
\end{minipage}
%=====================================
\centering
\scalebox{0.76}{
\begin{tabular}{|c|c|}
\hline 
\multicolumn{2}{|c|}{\T\B $\Op^{(P^3 V W_1)}$ terms  }  \\
\hline 
\T\B $ -4m_i^5\mathcal{I}_i^5 + 32m_i^3\mathcal{I}[q^2]_i^5 - 96m_i\mathcal{I}[q^4]_i^5  $ & $\epsilon^{\mu\nu\rho\sigma} P_{\mu} P_{\nu} P_{\rho} V_{\sigma} W_1 + \text{h.c.} $
    \\ \hline
\T\B $ -4m_i^5\mathcal{I}_i^5 + 32m_i^3\mathcal{I}[q^2]_i^5 - 96m_i\mathcal{I}[q^4]_i^5 $ & $\epsilon^{\mu\nu\rho\sigma} P_{\mu} P_{\nu} V_{\rho} P_{\sigma} W_1 + \text{h.c.} $
    \\ \hline
\end{tabular}
}
\caption{Subset operator structures in the degenerate fermionic UOLEA involving the pseudoscalar, vector and axial-vector structures. This subset will be used in the various examples we present in Sec.~\ref{sec:examples}.
\label{tab:VAoperatorexamples}}
\end{table}

\subsection{Results for the universal coefficients}
\label{sec:ResultsDescription}

We now present the results of the calculation outlined above, listing here only the UOLEA operators with $P$, $W_0$ and $W_1$ terms where all fermions in the loop are degenerate in mass. In this case, there are 52 distinct operator structures in the UOLEA, and we tabulate their coefficients in Tables~\ref{tab:pureP},~\ref{tab:mixedPW0},~\ref{tab:mixedPW1} and \ref{tab:mixedPW0W1}. The coefficients and operators containing only $P$'s can be found in Table~\ref{tab:pureP}. The operators contain the coupling with scalar structures $\Op^{(W_0)}, \, \Op^{(PW_0)}$ are tabulated in Table~\ref{tab:mixedPW0}, while the coupling with pseudo-scalar structures $\Op^{(W_1)}, \, \Op^{(PW_1)}$ are in Table~\ref{tab:mixedPW1}. Finally, the coefficients of the operators containing a mixture of scalar and pseudo-scalar structures $\Op^{(PW_0W_1)}$ are listed in Table~\ref{tab:mixedPW0W1}. Note that each universal coefficient in the Tables~\ref{tab:pureP},~\ref{tab:mixedPW0},~\ref{tab:mixedPW1} and \ref{tab:mixedPW0W1} has to be multiplied by the factor $i$, and that repeated Lorentz indices are implied to be contracted (though they are all written as subscripts for typographical convenience). 

Results for the more general non-degenerate mass spectrum and including the vector ($V$) and axial-vector ($A$) structures in the degenerate case are lengthy, so we include them in a Mathematica notebook made available on \texttt{GitHub} \href{https://github.com/HoaVuong-lpsc/The-Fermionic-UOLEA-Mathematica-notebook}{\faGithub}~\cite{notebookUOLEA}. Some of the UOLEA operators involving $V$ and $A$ that will be used in the examples in Sec.~\ref{sec:examples} are shown in Table~\ref{tab:VAoperatorexamples}.

For the user's convenience, we organised the Mathematica notebook as follows:
\begin{itemize}
    \item We remind the user that the effective Lagrangian will be a summation of all universal operators we have tabulated in the Mathematica notebook. The coefficient of each operator has to be multiplied by a factor of $i$. Afterward, we have to read off the value of the master integrals, as tabulated in Ref.~\cite{Zhang:2016pja}. We note that the coefficients include the $\mathcal{O}(\epsilon)$ terms that can cancel the $\frac{1}{\epsilon}$ pole from the loop integrals and yield finite contributions.
    \item In the first section of the Mathematica notebook, we summarise all universal structures as presented in the Tables~\ref{tab:W0W1P} and \ref{tab:Axial-only} where each entry is hyperlinked such that a click takes the user directly to the table of operator structures and their corresponding coefficients. 
    
    \item In the following sections, we present the full results in both degenerate and non-degenerate cases where the coupling matrix $X[\phi]$ contains only scalar and pseudo-scalar structures. 
    \item Finally, we present the full results in the degenerate case including the $V$ and $A$ structures. Due to a large number of combinations, we divide this section into subcategories: vector only, axial-vector only, and mixed vector/axial-vector. We also note that the results for mixed structures are written in functional space with open covariant derivatives. Depending on the effective operators one needs to construct, a subset of operators in the UOLEA will need to be selected and reorganized into the form of commutators. The non-denegerate results are available upon request.
    \item We use the same notation in the Mathematica notebook as in the Eq.~\eqref{eq:EFT-Lagrangian-tr} where \texttt{P},\texttt{W0}, \texttt{W1} stand for the covariant derivative, scalar, and pseudo-scalar structures, respectively. To avoid conflict with other Mathematica packages, we denote $\bar{v}^{b}$, $\bar{a}^{b}$ for vector and axial-vector structures. We follow the conventions of Ref.~\cite{Peskin:1995ev} for $\gamma^5$ and the total anti-symmetric tensor $\epsilon^{\mu\nu\rho\sigma}$, $\epsilon^{0123} = +1$. The trace of Dirac matrices is evaluated using the \texttt{FeynCalc} package~\cite{Mertig:1990an,Shtabovenko:2016sxi,Shtabovenko:2020gxv} and thus the output operator structures are also written in the language of this package. 
    \item Regarding the hermiticity of the operator structures, the operators which are not self-hermitian need to be accompanied with their hermitian conjugates. The non-self-hermitian operators appear with ``+ h.c." in the table of operators. We also checked that the operator and its hermitian conjugate have the same coefficients that result from the process of functional matching computations.  
\end{itemize}

\section{Examples} 
\label{sec:examples}

In this Section we present a few examples involving the top quark, as a cross-check of our results and to illustrate concretely how to use the fermionic UOLEA for practical calculations. 

\subsection{Integrating out the top quark in the Standard Model}

In the broken phase of the electroweak symmetry, the terms quadratic in the top quark field interacting with the SM Higgs via a Yukawa interaction are
\begin{align}
\mathcal{L}_{\rm SM} \supset \bar{t} \left( i\partial_{\mu} - g_sG_{\mu}^a T^a - e Q_t F_{\mu} \right)\gamma^{\mu} \, t - m_t\bar{t}t  - \dfrac{y_t}{\sqrt{2}}h \bar{t}t  \ ,
\label{eq:SM-Top-Yukawa}
\end{align}
where $G_\mu^a$ is the gluon field, $T^a$ is the $SU(3)_c$ generator, and $F_\mu$ is the notation chosen for the photon field so as to avoid confusion with the axial-vector matrix $A_\mu$.

The above Lagrangian can be written in the canonical form that provides the starting point for a UOLEA analysis as
\begin{align}
\mathcal{L}_{\rm SM}^{(\text{UOLEA form})} \supset \bar{t} \left( \gamma^{\mu}P_{\mu} - m_t - W_0   \right) t \ ,
\end{align}
where, for this example, the covariant derivative $P_{\mu}$ and the coupling matrix $W_0$ are
\begin{align}
P_{\mu} = iD_{\mu} = i\partial_{\mu} - g_sG_{\mu}^a T^a - e Q_t F_{\mu} \, , \quad W_0 = \dfrac{y_t}{\sqrt{2}}h.
\end{align}
We focus on the following operators in the EFT Lagrangian: $h$, $ h^2$, $(\partial_{\mu}h)^2$, $h\, F_{\mu\nu}F^{\mu\nu}$ and $h\, G^a_{\mu\nu}G^{a,\mu\nu}$. This selects the following relevant terms in the UOLEA:
\begin{align}
\mathcal{L}_{\rm EFT} \supset  -\dfrac{1}{(4 \pi )^2} &  \left[  \,  4m_t^3  \left( 1 -\log \dfrac{m_t^2}{\mu^2} \right)  \text{tr} W_0  + 2m_t^2 \, \left( 1 - 3 \log \frac{m_t^2}{\mu^2} \right) \text{tr} W_0^2  \right.
\nonumber  \\
& \quad  \left. - \left( \dfrac{2}{3} + \log \frac{m_t^2}{\mu^2} \right) \text{tr} [P_{\mu},W_0][P_{\mu},W_0] + \left(\dfrac{2}{3m_t}\right)\text{tr}\big( [P_{\mu},P_{\nu}][P_{\mu},P_{\nu}] W_0 \big) \, \right] ,
\label{eq:Example1-matchEqs1}
\end{align}
where the coefficient of each operator in Eqs.~\eqref{eq:Example1-matchEqs1} can be found in Table \ref{tab:mixedPW0}, and note that we must multiply those coefficients by $i$. To obtain the pre-computed coefficients in Eqs.~(\ref{eq:Example1-matchEqs1}), we must retain the $1/\epsilon$ poles in the master integrals. These poles can be multiplied by the $\epsilon$ terms appearing in the prefactor multiplying the master integral coming from the trace over gamma matrices in the operator.  
For example,
\begin{align}
(8-2\epsilon) \mathcal{I}[q^2]^2_i &= (8-2\epsilon) \dfrac{m_t^2}{2} \left( 1 - \log\dfrac{m_t^2}{\mu^2} + \dfrac{2}{\epsilon} -\gamma_E + \log 4\pi  \right)
\nonumber \\
&= 4m_t^2 \left( 1 - \log\dfrac{m_t^2}{\mu^2} \right) - 2m_t^2 \ ,
\end{align} 
where in going from the first to the second line, we take the limit $\epsilon \rightarrow 0$ and drop the terms $2/\epsilon -\gamma_E + \log 4\pi$, since we use the $\overline{\text{MS}}$-scheme for renormalisation. 

Next, we evaluate the trace over all internal indices, which in this case corresponds to the colour and $SU(3)_c$ indices carried by the top quark and gluon fields respectively, obtaining
\begin{align}
& \text{tr}W_0 = \text{tr} \, \dfrac{y_t}{\sqrt{2}} h \, \delta_{ab} = N_c \dfrac{y_t}{\sqrt{2}} h
\, , \quad
\text{tr}W_0^2 = \text{tr} \, \dfrac{y_t^2}{2} h^2 \, \delta_{ab}\delta_{ba} = N_c \dfrac{y_t^2}{2} h^2 \ ,
\nonumber \\
& \text{tr}[P_{\mu},W_0][P_{\mu},W_0] = \text{tr} \left[ i\partial_{\mu} - g_sG_{\mu}^a T^a - e Q_t F_{\mu} , \dfrac{y_t}{\sqrt{2} }h \, \delta_{ab}  \right] \left[ i\partial_{\mu} - g_sG_{\mu}^a T^a - e Q_t F_{\mu} , \dfrac{y_t}{\sqrt{2}} h \, \delta_{ba}  \right] 
\nonumber \\
&\qquad\qquad\qquad\quad ~~
= - N_c \dfrac{y_t^2}{2} (\partial_{\mu}h)^2 \ .
\label{eq:Example1-matchEqs2}
\end{align}
The field strength tensors can be obtained by using $\big[ P_{\mu},P_{\nu}\big] = i\left(-g_s G_{\mu\nu}^a T^a \right) + i\left( -eQ_t F_{\mu\nu} \right)$, 
\begin{align}
    \text{tr} \, \big( [P_{\mu},P_{\nu}][P_{\mu},P_{\nu}] W_0 \big)
    &= \text{tr} \bigg[\bigg(  (-ig_s)^2G^a_{\mu\nu}G^b_{\mu\nu}T^aT^b + (-ieQ_t)^2F_{\mu\nu}F_{\mu\nu} 
    \nonumber \\
    &\qquad\qquad\qquad\qquad\qquad\qquad 
    - 2(g_s \, eQ_t)G^a_{\mu\nu}T^a F_{\mu\nu} \bigg)  \left( \dfrac{y_t}{\sqrt{2}}h \, \delta^{cd} \right) \bigg]
    \nonumber \\
    &= -\left( N_c \, g_s^2 \dfrac{y_t}{2\sqrt{2}} \right) h G^a_{\mu\nu}G^a_{\mu\nu} - \left( N_c \, (eQ_t)^2 \dfrac{y_t}{\sqrt{2}} \right) h F_{\mu\nu}F_{\mu\nu} \ ,
\label{eq:Eample1-matchEqs3}
\end{align}
where $\text{tr} \left( T^a T^b\right) = \delta^{ab}/2$ for generators of the fundamental representation of an $SU(N)$ gauge group. Inserting \eqref{eq:Example1-matchEqs2},~\eqref{eq:Eample1-matchEqs3} into \eqref{eq:Example1-matchEqs1}, we obtain
\begin{align}
\mathcal{L}_{\rm EFT} &\supset  
\dfrac{-1}{(4\pi)^2} \left[ ~ y_t^2 \, N_c \left(\dfrac{1}{3}+\dfrac{1}{2}\log\dfrac{m_t^2}{\mu^2} \right)\left(\partial_{\mu}h \right)^2
\right.
\nonumber \\
&\qquad\qquad~\, \left. + \, 4\dfrac{y_t}{\sqrt{2}}N_c \, m_t^3 \left( 1-\log\dfrac{m_t^2}{\mu^2} \right)h 
+ y_t^2 \, N_c \, m_t^2 \left( 1-3\log\dfrac{m_t^2}{\mu^2} \right)h^2 \,
\right]
\nonumber \\
& + \left( \dfrac{y_t}{\sqrt{2}} \right) \left[ \, \dfrac{g_s^2}{48\pi^2  \, m_t} N_c\, hG^a_{\mu\nu}G^{a}_{\mu\nu} + \dfrac{e^2Q_t^2}{24\pi^2 \, m_t} N_c\, hF_{\mu\nu}F_{\mu\nu} \, \right] \ ,
\label{eq:Example1-Final}
\end{align}
where $N_c=3$, $Q_t=2/3$. The kinetic term for the Higgs may then be canonically normalised by a suitable field redefinition. The results of the first two lines of Eq.~(\ref{eq:Example1-Final}) agree with those of Ref.~\cite{Kramer:2019fwz}. The third line agrees with the results of Ref.~\cite{Gunion:1989we, Shifman:1979eb}.

\subsection{Integrating out the top quark coupling to a light pseudo-scalar Higgs $A^0$}
\subsubsection{The effective coupling $A^0\gamma\gamma$}
\label{sec:Agamgam}

In this example, we consider the top quark with a coupling to a light pseudo-scalar, denoted $A^0$. We assume this field is lighter than the top quark, so that we may integrate the latter out in order to obtain the Wilson coefficient for the dimension$-5$ operator coupling between $A^0$ and two photons. We assume a coupling structure of the pseudo-scalar to the top quark taking the same form as in the type II Two Higgs Doublet Model (2HDM) or the MSSM. Note, however, that our result may be generalised to any model involving a pseudo-scalar coupling to the top quark, by a simple rescaling.

The terms in the UV Lagrangian relevant for computing the effective $A^0\gamma\gamma$ coupling can be written in the form

%=====================
% 2HDM Lagrangian
%=====================
\begin{align}
\mathcal{L}_{\rm UV} &\supset \bar{t} \left[ \left(  i\partial_{\mu} - eQ_t F_{\mu}  \right)\gamma^{\mu} - m_t + i \dfrac{m_t}{v}\cot\beta A^0 \gamma^5  \right] t, 
\label{eq:MSSM-Lagrangian}
\end{align} 
where $g/2M_W=1/v$ and we use the notation of the 2HDM of type II,   $\tan\beta = v_1/v_2$ with $v = \sqrt{v_1^2 + v_2^2}$.\footnote{Note also the use of $F_\mu$ for the photon field, to avoid confusion with the axial-vector coupling matrix $A_\mu$.}

Upon integrating out the top-quark, we know that the effective interaction $A^0 \gamma\gamma$ should be of the form
\begin{align}
\mathcal{L}_{\rm EFT}\supset C_{A^0\gamma\gamma} \,  A^0 F_{\mu\nu}\tilde{F}^{\mu\nu} \ ,
\label{eq:EFT-Form}
\end{align}
where our convention for the dual field strength tensor is $\tilde{F}_{\mu\nu} \equiv \dfrac{1}{2}\epsilon_{\mu\nu\rho\sigma} F^{\rho\sigma}$, with $\epsilon^{0123} = +1$. The aim of this example is therefore to compute $C_{A^0\gamma\gamma}$ arising from the heavy top quark loop.

%================================================%
% Path Integral with UOLEA formula Approach
%================================================%
\subsubsection*{UV Lagrangian in the UOLEA form:}

Before using the pre-computed coefficients from the tables above, we need to write the UV Lagrangian in Eq.~\eqref{eq:MSSM-Lagrangian} in the UOLEA form in terms of the relevant structures, which in this case comprises of only $W_1$ in addition to the covariant derivative,
\begin{align}
\mathcal{L}_{\rm UV}^{(\text{UOLEA form})} \supset \bar{t} \left[ P_{\mu}\gamma^{\mu} - m_t - i\gamma^5 W_1  \right] t,
\label{eq:MSSM-uolea-form}
\end{align}
where the covariant derivative $P_{\mu}$ (omitting the gluon piece which does not contribute) and the coupling $W_1$ are
\begin{align}
P_{\mu} &\supset i\partial_{\mu} - eQ_t F_{\mu} \, ,  \quad W_1 = - \dfrac{m_t}{v}\cot \beta A^0  \ .
\end{align}
Clearly, the existence of only these two structures means we will only need operators and coefficients from Table~\ref{tab:mixedPW1} above in order to compute $C_{A^0\gamma\gamma}$. Furthermore, we know that since both $P_\mu$ and $W_1$ are dimension 1, we will need only operators from the table of dimension 5 to form the EFT Lagrangian operator. While in this example the power counting may seem superfluous since we are only interested in one operator with a transparent structure, in more complicated examples this counting can be extremely helpful.

\subsubsection*{Relevant structures in the UOLEA:}

Now, referring to Table~\ref{tab:mixedPW1}, we can immediately identify the necessary combinations of $P_\mu$ and $W_1$ that will form the effective operator $A^0 F_{\mu\nu}\tilde{F}^{\mu\nu} $, along with their universal coefficients (recalling that we must multiply the coefficients from the table by $i$). This therefore yields the effective Lagrangian as obtained from the UOLEA
\begin{align}
\mathcal{L}_{\rm EFT} &\supset i \left( \,  m_t^5\mathcal{I}_i^5 -  8m_t^3\mathcal{I}[q^2]^5_i + 24 m_t \mathcal{I}[q^4]^5_i  \right) \text{tr} \epsilon^{\mu\nu\rho\sigma} W_1 [P_{\mu},P_{\nu}] [P_{\rho},P_{\sigma}]
\nonumber \\
&= \dfrac{1}{32\pi^2 m_t} \text{tr} \left( \epsilon^{\mu\nu\rho\sigma} \, W_1 [P_{\mu},P_{\nu}] [P_{\rho},P_{\sigma}] \right)  \ .
\end{align}
The trace over the internal indices is then evaluated, to obtain
\begin{align}
\text{tr} \left( \epsilon^{\mu\nu\rho\sigma} W_1 [P_{\mu},P_{\nu}] [P_{\rho},P_{\sigma}] \right) &= \text{tr} \left( \left[ - \dfrac{m_t}{v}\cot\beta A^0 \right] (-ie Q_t)^2 \delta_{ab} \, \epsilon^{\mu\nu\rho\sigma} F_{\mu\nu} F_{\rho\sigma} \right)
\nonumber \\
&= 2 \, \dfrac{m_t}{v}\cot\beta (eQ_t)^2 N_c \, A^0 F_{\mu\nu}\tilde{F}_{\mu\nu}\ ,
\end{align}
where we have used that the commutator
$[P_{\mu},P_{\nu}] \phi =  i (- eQ_t) F_{\mu\nu} \phi$. Putting the two pieces together, we thus obtain the EFT operator corresponding to the effective interaction $A^0\gamma\gamma$,
\begin{align}
\mathcal{L}_{\rm EFT} %&=  \dfrac{1}{32\pi^2 m_t} \,  2 \dfrac{m_t}{v}\cot\beta (eQ_t)^2 N_c A F_{\mu\nu}\tilde{F}_{\mu\nu}
%\nonumber \\
&\supset \dfrac{e^2}{16 \pi^2 v}Q_t N_c \cot\beta  A^0 F_{\mu\nu}\tilde{F}_{\mu\nu} \ .
\label{eq:UOLEA-Matching-MSSM}
\end{align}
Comparing Eqs.~(\ref{eq:EFT-Form}) and (\ref{eq:UOLEA-Matching-MSSM}), we conclude that
\begin{equation}
C_{A^0\gamma\gamma} = \dfrac{e^2}{16 \pi^2 v}Q_t N_c \cot\beta \, .
\end{equation}
We have checked that this agrees with the result obtained by the usual Feynman diagram derivation. Eq.~(\ref{eq:UOLEA-Matching-MSSM}) also matches the one in Refs.~\cite{Kniehl:1995tn,  Spira:1995rr, Djouadi:2005gj} once the different convention for the dual field strength used in those references, $\tilde{F}_{\mu\nu} \equiv \epsilon_{\mu\nu\rho\sigma} F_{\rho\sigma}$, is taken into account. In contrast to the Feynman diagram computation, here the effective operator and its Wilson coefficient were trivially obtained using the pre-calculated universal results of the UOLEA and the simple evaluation of a trace over internal indices.

\subsubsection{The effective coupling $A^0ZZ$}

We next consider a more complicated matching procedure than in the previous examples. Indeed, since we wish to obtain the coefficient of the dimension-5 operator coupling the pseudo-scalar $A^0$ to $Z$ bosons, it is immediately apparent that we will now need to make use of the vector and axial-vector coupling matrices $V_\mu$ and $A_\mu$.

The relevant terms in the Lagrangian are 
\begin{align}
\mathcal{L}_{\rm UV} &\supset \bar{t} \left[ \left( i\partial_{\mu} \right)\gamma^{\mu} - m_t + \left( i \dfrac{m_t}{v} \cot\beta A^0 \right) \gamma^5 \right.
\nonumber \\
& \qquad\qquad\qquad \left. - \dfrac{g}{\cos \theta_w} \left( \dfrac{T_3}{2} - Q_t \sin^2\theta  \right)Z_{\mu} \gamma^{\mu} + \left( \dfrac{g}{\cos\theta_w} \dfrac{T_3}{2} \right)Z_{\mu} \gamma^{\mu}\gamma^5  \right] t \, ,
\label{eq:MSSM-AZZ-Lagrangian}
\end{align}
where we used the same conventions as in Ref.\cite{Djouadi:2005gj, Gunion:1991cw}. Meanwhile, the effective Lagrangian for the $A^0ZZ$ effective coupling is
\begin{align}
\mathcal{L}_{\rm EFT} \supset C_{A^0ZZ} A^0 Z_{\mu\nu} \tilde{Z}^{\mu\nu},
\label{eq:EFT-AZZ-Lagrangian}
\end{align}
where $\tilde{Z}^{\mu\nu} = \dfrac{1}{2} \epsilon^{\mu\nu\rho\sigma} \partial_{[\rho} Z_{\sigma]}$, with $\epsilon^{0123} = +1$. 

%===========================================
% UOLEA
%===========================================

\subsubsection*{UV Lagrangian in the UOLEA form:} 

As in the previous examples, we first re-write the UV Lagrangian in terms of the UOLEA structures,
\begin{align}
\mathcal{L}_{\rm UV}^{(\text{UOLEA form})} \supset \bar{t} \left[ P_{\mu}\gamma^{\mu} - m_t - i \gamma^5 W_1 - V_{\mu}\gamma^{\mu} - A_{\mu}\gamma^{\mu}\gamma^5  \right] t,
\end{align}
where the $P_{\mu}, \, W_1, \, V_{\mu}, \, A_{\mu}$ are defined as
\begin{align}
\begin{cases}
P_{\mu} &\supset i\partial_{\mu} \, , 
\\
W_1 &= - \dfrac{m_t}{v} \cot\beta A^0 \ ,
\\
V_{\mu} &= g_V Z_{\mu} \, ; \quad g_V = \dfrac{g}{\cos \theta_w} \left( \dfrac{T_3}{2} - Q_t \sin^2\theta_w  \right)  \ ,
\\ 
A_{\mu} &= g_A Z_{\mu} \, ; \quad g_A = - \left( \dfrac{g}{\cos\theta_w} \dfrac{T_3}{2} \right) \, .
\end{cases}
\end{align}
We have dropped the gluon and photon pieces in the covariant derivative, since they do not contribute to the matching calculation here.

\subsubsection*{Relevant structures in the UOLEA:}

Having identified the UOLEA structures that will appear in the construction of the EFT operator, we must now decompose the latter to determine what UOLEA structures will contribute. The EFT Lagrangian is

\begin{align}
\mathcal{L}_{\rm EFT} &\supset C_{A^0ZZ} A^0 Z_{\mu\nu} \tilde{Z}^{\mu\nu} 
= \dfrac{1}{2} C_{AZZ} A^0 \epsilon^{\mu\nu\rho\sigma} \left( \partial_{\mu}Z_{\nu} - \partial_{\nu}Z_{\mu}  \right) \left( \partial_{\rho}Z_{\sigma} - \partial_{\sigma}Z_{\rho} \right)
\label{eq:analyze-EFT-AZZ}
\end{align}
Thus, to reconstruct the EFT operator in terms of UOLEA structures, we need
\begin{itemize}
\item One insertion of $W_1$ to account for the pseudo-scalar $A^0$.
\item Two $P_{\mu}$ insertions to account for the partial derivatives.
\item To account for the two $Z$ bosons, one might expect some combination of the structures $AA$, $VV$ or $AV$ to be required. However, due to the structure of the effective operator, it is clear that the product of the various UOLEA coupling matrices should have an odd number of $\gamma^5$ insertions. Since $W_1$ carries a $\gamma^5$, $AV$, which also has one $\gamma^5$, will not contribute.
\end{itemize}
The EFT Lagrangian will therefore be given by the following classes of UOLEA operators, 
\begin{align}
\mathcal{L}_{\rm EFT} \supset \sum_N\, f_N \Op_N^{\left( P^2 V^2 W_1 \right)} + f_N \Op_N^{\left( P^2 A^2 W_1 \right)} \, .
\label{eq:UOLEA-AZZ-MainTerms}
\end{align} 
Note that since we are integrating out the top-quark coupling to a single type of vector boson, the $W_1, \, V_{\mu}, \, A_{\mu}$ terms are proportional to the identity matrix and so commute with each other, which will simplify the calculation. Owing to the commutativity of $W_1$ with $A_\mu$ and $V_\mu$, the combination of all the $\Op^{(P^2A^2W_1)}$ and $\Op^{(P^2V^2W_1)}$ UOLEA operators of Table~\ref{tab:VAoperatorexamples} can be written as
\begin{align}
\mathcal{L}_{\rm EFT} &\supset f_1 \text{tr}\left( \epsilon^{\mu\nu\rho\sigma}P_{\mu}P_{\nu}V_{\rho}V_{\sigma} W_1  \right) + f_2 \text{tr}\left( \epsilon^{\mu\nu\rho\sigma}P_{\mu} W_1 P_{\nu}V_{\rho}V_{\sigma}  \right)
\nonumber \\
& \quad + f_3 \text{tr}\left( \epsilon^{\mu\nu\rho\sigma}P_{\mu}P_{\nu}A_{\rho}A_{\sigma} W_1  \right) + f_4 \text{tr}\left( \epsilon^{\mu\nu\rho\sigma}P_{\mu} W_1 P_{\nu}A_{\rho}A_{\sigma}  \right), 
\end{align}
where the values of the universal coefficients are
\begin{align}
f_1 &= i \left( 4 m_i^5 \mathcal{I}^5_i - 32m_i^3 \mathcal{I}[q^2]^5_i + 96m_i \mathcal{I}[q^4]^5_i \right) =  \dfrac{1}{8\pi^2 m_t}  \ ,
\nonumber \\
f_2 &=  i \left( -4 m_i^5 \mathcal{I}^5_i + 32m_i^3 \mathcal{I}[q^2]^5_i - 96m_i \mathcal{I}[q^4]^5_i \right) = \dfrac{-1}{8\pi^2 m_t}  \ ,
\nonumber \\
f_3 &=  i \left( -4 m_i^5 \mathcal{I}^5_i + 96m_i \mathcal{I}[q^4]^5_i \right)  =  \dfrac{1}{24\pi^2 m_t}  \ ,
\nonumber \\
f_4 &=  i \left( 4 m_i^5 \mathcal{I}^5_i - 96m_i \mathcal{I}[q^4]^5_i \right) = \dfrac{-1}{24\pi^2 m_t}  \ .
\end{align}

These UOLEA operators and their coefficients can be read off from Table~\ref{tab:VAoperatorexamples}; a complete tabulation of UOLEA results for the degenerate vector and axial-vector case can be found in the accompanying Mathematica notebook, as described in Sec.~\ref{sec:ResultsDescription}. 
Due to the proliferation of UOLEA operators involving $V$ and $A$, these are not listed in a commutator basis. Instead, it is preferable to perform the rearrangement into the commutator basis for the small subset of operators contributing to a specific application. We will now demonstrate this for the effective $A^0ZZ$ coupling.

\subsubsection*{Constructing the EFT operators:} 
We begin with the vector structure case, 
\begin{align}
\mathcal{L}_{\rm EFT}^{(\text{vector})} &=  f_1  \text{tr} \, \epsilon^{\mu\nu\rho\sigma} P_{\mu} P_{\nu} V_{\rho} V_{\sigma} W_1  + f_2 \text{tr} \, \epsilon^{\mu\nu\rho\sigma} P_{\mu} W_1 \, P_{\nu} V_{\rho} V_{\sigma}  \, ,
\label{eq:AZZ-UOLEA-CDbasis}
\end{align}
and rearrange it into the following commutator basis (note that e.g. $[V, \, W_1]=0$),
\begin{align}
\mathcal{L}_{\rm EFT}^{(\text{vector})} &\supset c_1  \text{tr} \, \epsilon^{\mu\nu\rho\sigma} W_1 \left[ P_{\mu}, V_{\nu} \right] \left[ P_{\rho}, V_{\sigma} \right]
+ c_2 \text{tr} \, \epsilon^{\mu\nu\rho\sigma} W_1 \left[ P_{\mu}, P_{\nu} \right] \left[ V_{\rho}, V_{\sigma} \right]
\nonumber \\
&= -c_1 \text{tr} \, \epsilon^{\mu\nu\rho\sigma} P_{\mu}W_1 P_{\nu} V_{\rho}V_{\sigma} + (4c_2 -c_1) \text{tr} \, \epsilon^{\mu\nu\rho\sigma} P_{\mu}P_{\nu}V_{\rho}V_{\sigma} W_1 \, .
\label{eq:AZZ-UOLEA-CCbasis}
\end{align}
In the second line we have expanded the commutators, so that comparing with the non-commutator basis of Eqs.\eqref{eq:AZZ-UOLEA-CDbasis} then allows us to solve a system of linear equations that take us from the non-commutator to the commutator basis of operators, 
\begin{align}
\begin{cases}
-c_1 = f_2 
\\
4c_2 - c_1 = f_1
\end{cases}
\Leftrightarrow
\begin{cases}
c_1 = -f_2 =  \dfrac{1}{8\pi^2 m_t}
\\ 
c_2 = \dfrac{f_1 - f_2}{4} =  \dfrac{1}{16 \pi^2 m_t}
\end{cases}
\end{align}
Using $[P_{\mu},V_{\nu}] = i( \partial_{\mu} V_{\nu} )$, we may rewrite the operator $\text{tr} \, \epsilon^{\mu\nu\rho\sigma} \, W_1  [P_{\mu}, V_{\nu}] [P_{\rho},V_{\sigma}]$ as 
\begin{align}
c_1 \, \text{tr} \, \epsilon^{\mu\nu\rho\sigma} \, W_1  [P_{\mu}, V_{\nu}] [P_{\rho},V_{\sigma}] 
&= (i^2) \, c_1 \, \text{tr} \, \epsilon^{\mu\nu\rho\sigma} W_1 (\partial_{\mu}V_{\nu})(\partial_{\rho}V_{\sigma})
\nonumber \\
&= c_1 \dfrac{i^2}{4} \text{tr} \, W_1 \left[ \epsilon^{\mu\nu\rho\sigma}  (\partial_{\mu}V_{\nu})(\partial_{\rho}V_{\sigma}) + \epsilon^{\mu\nu\sigma\rho}  (\partial_{\mu}V_{\nu})(\partial_{\sigma}V_{\rho}) \right.
\nonumber \\
&\qquad\qquad\quad + \epsilon^{\nu\mu\rho\sigma}  (\partial_{\nu}V_{\mu})(\partial_{\rho}V_{\sigma}) + \epsilon^{\nu\mu\sigma\rho}  (\partial_{\nu}V_{\mu})(\partial_{\sigma}V_{\rho}) \,  ]
\nonumber \\
&= c_1 \dfrac{i^2}{4} \text{tr} \, \epsilon^{\mu\nu\rho\sigma} W_1 [\partial_{\mu}V_{\nu} - \partial_{\nu}V_{\mu} ] [\partial_{\rho}V_{\sigma} - \partial_{\sigma}V_{\rho} ].
\end{align}
Putting it all together, we obtain the contributions from the vector terms $\Op^{(P^2V^2W_1)}$,
\begin{align}
\mathcal{L}_{\rm EFT}^{(\text{vector})} &\supset \dfrac{1}{8\pi^2 m_t} \dfrac{i^2}{4} \text{tr} \, \epsilon^{\mu\nu\rho\sigma} \left( -\dfrac{m_t}{v}\cot\beta A^0 \right)\delta_{ab} \, g_V^2 Z_{\mu\nu}Z_{\rho\sigma}
\nonumber \\
&= \dfrac{1}{16 \pi^2 v} N_c \cot\beta \dfrac{g^2}{\cos^2 \theta_w} \left( \dfrac{T_3}{2} - Q_t \sin^2\theta_w \right)^2 A^0 Z_{\mu\nu}\tilde{Z}_{\mu\nu} \, .
\label{eq:AZZ-vector}
\end{align}
The computation for the UOLEA operators involving the axial-vector coupling matrix $A$ proceeds similarly. We find
\begin{align}
\mathcal{L}_{\rm EFT}^{(\text{axial-vector})} &\supset
\dfrac{1}{24\pi^2 m_t} \dfrac{i^2}{4} \text{tr} \, \epsilon^{\mu\nu\rho\sigma} \left( -\dfrac{m_t}{v}\cot\beta A^0 \right)\delta_{ab} \, g_A^2 Z_{\mu\nu}Z_{\rho\sigma}
\nonumber \\
&= \dfrac{1}{48 \pi^2 v} N_c \cot\beta \left( \dfrac{g}{\cos\theta_w} \dfrac{T_3}{2} \right)^2 A^0 Z_{\mu\nu}\tilde{Z}_{\mu\nu}. 
\label{eq:AZZ-axial}
\end{align}
Adding (\ref{eq:AZZ-vector}) and (\ref{eq:AZZ-axial}) gives the final result, 
\begin{align}
\mathcal{L}_{\rm EFT} \supset \dfrac{1}{48 \pi^2 v} N_c \cot\beta \dfrac{g^2}{\cos^2 \theta_w} \left( T_3^2 + 3Q_t \sin^2\theta_w [ Q_t \sin^2 \theta_w - T_3] \right) A^0 Z_{\mu\nu}\tilde{Z}_{\mu\nu}.
\end{align}
This result agrees with the one in Ref.\cite{Kniehl:1995tn}. However, the calculation here is carried out in a more streamlined manner using the UOLEA.

\subsubsection{The effective coupling $A^0 Z\gamma$}
To construct the effective coupling $A^0 Z\gamma$ resulting from integrating out the top quark coupling to a light pseudo-scalar $A^0$, we split the interaction with the $Z$ boson into vector and axial-vector currents. The relevant terms in the UV Lagrangian are then
\begin{align}
\mathcal{L}_{\rm UV} &\supset \bar{t} \left[ \left( i\partial_{\mu} - eQ_t F_{\mu} \right)\gamma^{\mu} - m_t + \left( i \dfrac{m_t}{v} \cot\beta A^0 \right) \gamma^5 \right.
\nonumber \\
& \qquad\quad\quad \left. - \dfrac{g}{\cos \theta_w} \left( \dfrac{T_3}{2} - Q_t \sin^2\theta_w  \right)Z_{\mu} \gamma^{\mu} + \left( \dfrac{g}{\cos\theta_w} \dfrac{T_3}{2} \right)Z_{\mu} \gamma^{\mu}\gamma^5  \right] t \, ,
\label{eq:MSSM-AZg-Lagrangian}
\end{align}
where $F_{\mu}$ denotes the photon field. We now integrate out the top quark to obtain the following $\mathcal{C}\mathcal{P}$-even effective operator,
\begin{align}
\mathcal{L}_{\rm EFT} \supset C_{A^0Z\gamma} A^0 Z_{\mu\nu} \tilde{F}^{\mu\nu}.
\label{eq:AZg-EFT-OP}
\end{align}

%=================================
% uolea
%=================================
\subsubsection*{UV Lagrangian in the UOLEA form:} 

We write the UV Lagrangian (\ref{eq:MSSM-AZg-Lagrangian}) in the canonical form,
\begin{align}
\mathcal{L}_{\rm UV}(\text{UOLEA form}) = \bar{t} \left[ P_{\mu}\gamma^{\mu} - m_t - i \gamma^5 W_1 - V_{\mu}\gamma^{\mu} - A_{\mu}\gamma^{\mu}\gamma^5 \right] t,
\end{align}
where the structures $P_{\mu}, W_1, V_{\mu}, A_{\mu}$ correspond to
\begin{align}
\begin{cases}
P_{\mu} &\supset i\partial_{\mu} - eQ_tF_{\mu} \, , 
\\
W_1 &= - \dfrac{m_t}{v} \cot\beta A^0 \ ,
\\
V_{\mu} &= g_V Z_{\mu} \, ; \quad g_V = \dfrac{g}{\cos \theta_w} \left( \dfrac{T_3}{2} - Q_t \sin^2\theta_w  \right)  \ ,
\\ 
A_{\mu} &= g_A Z_{\mu} \, ; \quad g_A = - \left( \dfrac{g}{\cos\theta_w} \dfrac{T_3}{2} \right) \, .
\end{cases}
\label{eq:fooPVWA}
\end{align}
Note that after the broken phase, our theory still respects $U(1)_\text{QED}$, thus the photon field still lives in the covariant derivative (together with the gluon field, which does not contribute in the present case and has been omitted), while the $Z$ boson should be put into the $V$ and $A$ structures.

\subsubsection*{Relevant structures in the UOLEA:}
To obtain the EFT operator (\ref{eq:AZg-EFT-OP}), we need: 
\begin{itemize}
\item One insertion of $W_1$ to account for the appearance of $A^0$.
\item Three insertions of $P_{\mu}$. Two of them form the photon field strength. The last one will act on the $Z_{\mu}$. Then combining with the anti-symmetric tensor $\epsilon^{\mu\nu\rho\sigma}$ we can construct the dual field-strength tensor of the $Z$ boson. 
\item One insertion of $V$ to account for $Z_\mu$. As in the previous example, we can count $\gamma^5$ insertions to see that no operator involving $A$ can contribute to the EFT operator \eqref{eq:AZg-EFT-OP}. 
\end{itemize}
Putting it all together, the relevant class of UOLEA operators which contribute to the EFT operator~\eqref{eq:AZg-EFT-OP} is then
\begin{align}
\mathcal{L}_{\rm EFT}^{(\rm UOLEA)} \supset \sum_N f_N \Op_N^{\left( P^3 V W_1 \right)} \, .
\end{align}
Since in this case $[W_1, V_{\mu}] = 0$, we have only one UOLEA operator to consider,
\begin{align}\label{eq:AZg-P3VW1-cd}
\mathcal{L}_{\rm EFT}^{(\text{vector})} = i \left( -4m_i^5\mathcal{I}_i^5 + 32m_i^3\mathcal{I}[q^2]_i^5 - 96m_i\mathcal{I}[q^4]_i^5 \right) \bigg[ \text{tr}\left( \epsilon^{\mu\nu\rho\sigma}P_{\mu}P_{\nu}V_{\rho}P_{\sigma}W_1 \right) + \text{h.c.} \bigg] \, , 
\end{align}
we note that the operator structure $\big[  \text{tr}\left( \epsilon^{\mu\nu\rho\sigma}P_{\mu}P_{\nu}P_{\rho}V_{\sigma}W_1 \right) + \text{h.c.}  \big]$ vanishes due to the antisymmetry of the $\epsilon^{\mu\nu\rho\sigma}$ tensor and $[W_1, V_{\mu}] = 0$. We then rearrange the operator structures in \eqref{eq:AZg-P3VW1-cd} into the basis where $P$'s only appear in the commutators. The operator structures we expect in the commutator basis are
\begin{align}\label{eq:AZg-P3VW1-commute}
    \mathcal{L}_{\rm EFT}^{(\text{vector})} &\supset f_1 \big(\, \text{tr}\, \epsilon^{\mu\nu\rho\sigma}\big[P_{\mu},P_{\nu}\big] \big[P_{\rho},V_{\sigma}\big]W_1 + \text{tr}\,\epsilon^{\mu\nu\rho\sigma}W_1\big[P_{\mu},V_{\nu}\big] \big[P_{\rho},P_{\sigma} \big] \, \big)
    \nonumber\\
    &= 2f_1 \big(\, \text{tr}\, \epsilon^{\mu\nu\rho\sigma}P_{\mu}P_{\nu}V_{\rho}P_{\sigma}W_1 + \text{tr}\, \epsilon^{\mu\nu\rho\sigma}P_{\mu}P_{\nu}W_1 P_{\rho}V_{\sigma}  \, \big)
\end{align}
As in the previous examples, we expand the commutators and, using the fact that $\left[W_1,V_{\mu} \right]=0$, match with the non-commutator basis of Eq.~\eqref{eq:AZg-P3VW1-cd} and fix the value of the coefficient $f_1$:
\begin{align}
    f_1 = i \, \dfrac{1}{2} \left( -4m_i^5\mathcal{I}_i^5 + 32m_i^3\mathcal{I}[q^2]_i^5 - 96m_i\mathcal{I}[q^4]_i^5 \right)
    = \dfrac{-1}{16 \pi^2 m_t} \, .
\end{align}
Plugging $P_{\mu}, \, V_{\mu}$ and $W_1$ from Eq.~\eqref{eq:fooPVWA} into
\begin{align}
    \mathcal{L}_{\rm EFT}^{(\text{vector})} &\supset f_1 \big(\, \text{tr}\, \epsilon^{\mu\nu\rho\sigma}\big[P_{\mu},P_{\nu}\big] \big[P_{\rho},V_{\sigma}\big]W_1 + \text{tr}\,\epsilon^{\mu\nu\rho\sigma}W_1\big[P_{\mu},V_{\nu}\big] \big[P_{\rho},P_{\sigma} \big] \, \big) \, ,
\end{align}
and using $\left[ P_{\mu},P_{\nu} \right] = i\left(-eQ_t \right)F_{\mu\nu}$ and $\left[ P_{\mu},V_{\nu} \right] = i g_V \left( \partial_{\mu}Z_{\nu} \right)$, we obtain
\begin{align}
\mathcal{L}_{\rm EFT} &\supset f_1 \left(g_V \, eQ_t \right) \text{tr}\,\big[ 2 \epsilon^{\mu\nu\rho\sigma} \left(\partial_{\mu}Z_{\nu}\right) F_{\rho\sigma}W_1 \big]
\nonumber \\
&= f_1 \left(g_V \, eQ_t \right) \text{tr}\,\big[ \epsilon^{\mu\nu\rho\sigma} \left(\partial_{\mu}Z_{\nu}\right) F_{\rho\sigma}W_1 + \epsilon^{\nu\mu\rho\sigma} \left(\partial_{\nu}Z_{\mu}\right) F_{\rho\sigma}W_1 \big]
\nonumber \\
&= \dfrac{-1}{16 \pi^2 m_t} \left(g_V \, eQ_t \right) \text{tr} \left[ \left(-\dfrac{m_t}{v}\cot\beta A^0 \, \delta^{ab} \right) \epsilon^{\mu\nu\rho\sigma} Z_{\mu\nu}F_{\rho\sigma} \right] \ .
\end{align}
Taking the trace over colour degrees of freedom and using $g_V = \dfrac{g}{\cos \theta_w} \left(\dfrac{T_3}{2} - Q_t\sin^2\theta_w \right)$ from Eq.~\eqref{eq:fooPVWA}, we obtain the final result, 
\begin{align}
\mathcal{L}_{\rm EFT}
&\supset \dfrac{1}{16\pi^2 v} N_c \cot\beta ( eQ_t ) \dfrac{g}{\cos\theta_w}\left( T_3 - 2Q_t \sin^2\theta_w \right) A^0 Z_{\mu\nu}\tilde{F}_{\mu\nu} \ .
\label{eq:AZg-UOLEA-Lagrangian}
\end{align}
This result agrees with the ones in Refs.\cite{Gunion:1989we, Kniehl:1995tn}. Once again we note the relative ease and efficiency with which the same result can be derived in the UOLEA.

\section{Conclusion}
\label{sec:conclusion}

The universality of the one-loop effective action obtained by integrating out heavy degrees of freedom has emerged as a byproduct of improved path integral methods for performing these calculations. This so-called UOLEA makes the repeated evaluation of functional determinants redundant and provides a more efficient way of matching at one loop compared to Feynman diagrams, especially when systematically obtaining an ensemble of operator coefficients at once. It also has the advantage of being easier to automate. 

Previous work developed the bosonic UOLEA for integrating out heavy bosons, including mixed heavy-light loops. While these results could be used for integrating out fermions as well in some cases, they did not account for $\gamma$ matrices in the fermion couplings, and were also not as straightforward to use as in the bosonic case. It was therefore necessary to extend the UOLEA to the fermionic case, and desirable to do so in a way that maintained the simplicity of the UOLEA approach. 

In this work we presented the fermionic UOLEA, which can be used for one-loop matching with heavy fermions in the loop, coupling with structures involving $\gamma$ matrices. The starting point is the UV Lagrangian of Eq.~\eqref{eq:LUVfermionic}, for which the UOLEA is given by Eq.~\eqref{eq:UOLEA_fermionic}. A subset of our results for the new UOLEA operators and the corresponding universal coefficients are tabulated in Tables~\ref{tab:pureP},~\ref{tab:mixedPW0},~\ref{tab:mixedPW1}, \ref{tab:mixedPW0W1} and~\ref{tab:VAoperatorexamples} for the degenerate mass case, while the full results, in the non-degenerate case for $P, W_0, W_1$ structures and in the degenerate case for $V, A$ structures, are available in the accompanying Mathematica notebook~\href{https://github.com/HoaVuong-lpsc/The-Fermionic-UOLEA-Mathematica-notebook}{\faGithub},~\cite{notebookUOLEA}.\footnote{The non-degenerate results for $V, A$ structures can also be made available on request.} These expressions can be readily incorporated into codes that automate the tracing over the internal indices and the rearranging of the resulting EFT operators into a non-redundant basis.\footnote{See, for example, Ref.~\cite{Criado} for automated tree-level matching and Ref.~\cite{Codex} that implements the degenerate bosonic UOLEA results.}   

The status of the UOLEA terms available and those that remain to be computed is summarised in Table~\ref{tab:UOLEAprogress}. 
This is listed for completeness though we note that the majority of UV Lagrangian structures of interest are now included in the UOLEA for obtaining EFT operators up to dimension 6. 
Nevertheless, further efforts to complete the UOLEA, including all possible structures and extending to higher dimensional operators, would then enable and be a part of a fully general automated one-loop matching tool. This ambitious goal is left for future work.

\subsubsection*{Acknowledgments} 

We thank the authors of Ref.~\cite{AngelescuPeizi} for making us aware of their related work on fermionic results for the UOLEA being finalised at the same time as our manuscript. J.Q. thank the authors of Ref.~\cite{Kramer:2019fwz} for useful discussions. T.Y. is supported by a Branco Weiss Society in Science fellowship and partially supported by STFC consolidated grant ST/P000681/1. S.A.R.E.\ would like to thank the ``Institut de Physique Th\'eorique de La C\^ote" for hospitality while this work was completed.
S.A.R.E.\ is supported by the U.S.\ Department of Energy under Contract No.\ DE-AC02-76SF00515 and by the Swiss National Science Foundation, SNF project number P400P2$\_$186678.
Z.Z.'s work is supported by the U.S.\ Department of Energy, Office of Science, Office of High Energy Physics, under Award Number DE-AC02-05CH11231.

\bibliographystyle{JHEP}
\bibliography{biblio}

\end{document}